\documentclass[conference]{IEEEtran}

\usepackage{cite}
\usepackage{graphicx}
\usepackage{lipsum}

\usepackage{subcaption}
\usepackage{refstyle}
\usepackage[cmex10]{amsmath}
\usepackage{mathtools}
\DeclarePairedDelimiter\ceil{\lceil}{\rceil}
\DeclarePairedDelimiter\floor{\lfloor}{\rfloor}
\usepackage[ruled,vlined]{algorithm2e}
\include{pythonlisting}
\usepackage{array}
\usepackage{url}
\usepackage{fontenc}
\usepackage[utf8]{inputenc}
\usepackage{mathtools}
\usepackage{tabularx}
\usepackage{tabu}
\usepackage{float}
\usepackage{multirow}
\usepackage{siunitx}
\usepackage{algorithmic}
\usepackage{psfrag}
\usepackage{pgfplots}
\usepackage{xcolor}
\newcommand\note[1]{\textcolor{red}{#1}}
\usepackage{comment}
\usepackage{wrapfig}

\usepackage[utf8]{inputenc}
\usepackage{amsmath,amssymb}
\usepackage{enumitem}

\newsavebox{\ieeealgbox}

\newcolumntype{M}[1]{>{\arraybackslash}m{#1}}
\newcolumntype{N}{@{}m{0pt}@{}}

\usepackage{proof}
\usepackage{breqn}

\begin{document}



\title{Learning Secured Modulation With \\Deep Adversarial Neural Networks}

\author{\IEEEauthorblockN{Hesham Mohammed and Dola Saha}
\IEEEauthorblockA{\textit{Department of Electrical \& Computer Engineering} \\
\textit{University at Albany, SUNY}\\
\{hhussien, dsaha\} @albany.edu}
}

\IEEEoverridecommandlockouts
\maketitle
\pagestyle{plain}


\begin{abstract}

Growing interest in utilizing the wireless spectrum by heterogeneous devices compels us to rethink the physical layer security to protect the transmitted waveform from an eavesdropper. 
We propose an end-to-end symmetric key neural encryption and decryption algorithm with a modulation technique, which remains undeciphered by an eavesdropper, equipped with the same neural network and trained on the same dataset as the intended users. We solve encryption and modulation as a joint problem for which we map the bits to complex analog signals, without adhering to any particular encryption algorithm or modulation technique. We train to cooperatively learn encryption and decryption algorithms between our trusted pair of neural networks, while eavesdropper's model is trained adversarially on the same data to minimize the error. We introduce a discrete activation layer with a defined gradient to combat noise in a lossy channel. Our results show that a trusted pair of users can exchange data bits in both clean and noisy channels, where a trained adversary cannot decipher the data.

\end{abstract}

\section{Introduction}

As new spectrum (sub-6GHz, mmWave and TeraHertz) becomes available for communication and coexistence of frequency-agile cognitive heterogeneous nodes becomes a norm, we need to rethink physical layer security to provide maximum secrecy of the waveforms in a broadcast channel. Recent advances in the use of neural networks for communication systems intrigue us to investigate whether neural networks can be trained to simultaneously learn an encryption/decryption algorithm as well as modulate/demodulate bits to transmit and receive an analog signal. Neural networks are applied to accomplish complex tasks with end-to-end data based training in order to achieve a certain objective or minimize a certain loss function. These tasks can be generating images~\cite{gregor2015draw}, learning complex distributions, classification~\cite{ou2007multi} or performing autonomous driving~\cite{sallab2017deep}. On the other hand, most of the cryptography algorithm operates on bit manipulation, whereas neural networks work on continuous signals. Instead of combating these discrepancies between neural networks and cryptography, we leverage the efficiency of neural network in solving complex tasks and its capability of handling continuous signal to our advantage in bridging the gap between cryptography and modulation. At the same time, we are broadening the scope of cryptography beyond the bits domain and mapping to a much larger complex domain. This ensures that it becomes computationally more complex to decrypt using Brute Force attacks. Although Physical layer security research ~\cite{Poor19, Survey, sec_cogradio_13, sec_cogradio_15} 
takes advantage of the channel impairments to modify the transmitted signal, it does not ensure a complete optimized system that can take in bits and convert them to secured waveforms.  


Learning an encryption and decryption algorithm is beyond just mapping of bits to another domain. To ensure secrecy and integrity of information, it is important that an eavesdropper (Eve) listening to the cipher data (communication between Alice, the sender, and Bob, the receiver) will not be able to decipher it without a key. Hence, we design Alice and Bob to each have a neural network to learn the optimized algorithm. 
This notion is same as the encryption and decryption algorithms are known is classical cryptography.
It is only possible to obtain higher levels of secrecy when we can also emulate an adversary, a passive eavesdropper in our model, and cooperatively train Alice and Bob to defeat Eve. Hence, we introduce Eve as another neural network during the training phase that is constantly trying to decipher the data and the model is trained in an adversarial manner to converge to a system, where Alice and Bob can defeat Eve. We also train Eve on the same model and with the same plain text data as Alice and Bob. This guarantees that the learned encryption algorithm will be able to beat both a trained as well as untrained Eve with same neural network structure as Alice and Bob. This paper aims at combining the cryptography and modulation using neural networks, which can be extended later to implement physical layer security in OFDM and massive MIMO systems.

In this paper, three neural networks are trained simultaneously to learn a symmetric key based secured modulation method in presence of a passive eavesdropper. The key contributions of this work can be listed as follows:
\newcounter{contrib}

\stepcounter{contrib}
\noindent
\arabic{contrib}) We propose an end-to-end learning of shared key based secured modulation technique, where bits are mapped to real numbers and then converted to complex domain to be transmitted over a channel.\\
\stepcounter{contrib}
\noindent
\arabic{contrib}) We introduce a discrete activation layer with a defined gradient, which is derived to support a practical lossy medium communication system and finite memory devices. The activation function guarantees a gradient, when stochastic gradient descent (SGD)~\cite{SGD} is applied during training.\\
\stepcounter{contrib}
\noindent
\arabic{contrib}) We design our system so that it is able to adapt to both clear and noisy channels to ensure that secured communication can be carried out between trusted parties while a passive eavesdropper is unable to decipher it.

\begin{figure*}
    \centering
    \includegraphics[width=0.7\linewidth]{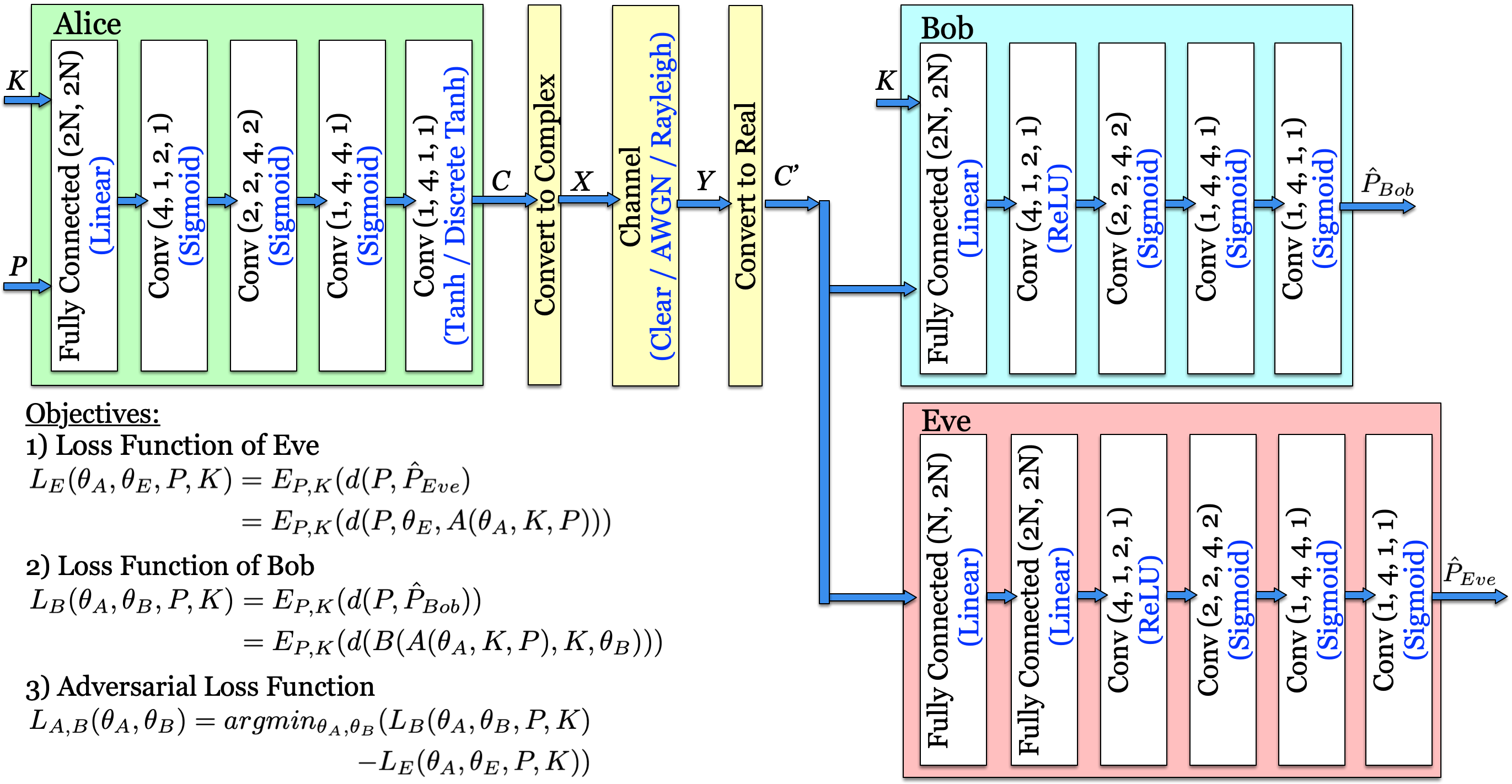}
    \caption{Neural Network Architecture}
    \label{fig:nn}
\end{figure*}

\section{Related Work}
\label{relared_work}
Deep neural network has been used in physical layer communication~\cite{AE} to learn optimal constellation points without prior mathematical formulation. Also, GANs~\cite{AE_GANS} have been used to train the autoencoder network for practical channel models. In addition, autoencoders are used to learn advanced communication schemes, such as orthogonal frequency division multiplexing (OFDM), which enables reliable transmission in wireless channels~\cite{AE_ofdm} as well as optical media~\cite{H_AE}.

The idea of using adversarial neural network in cryptography was introduced in~\cite{abadi2016learning}. 
However, the authors restrict the model to work with floating point domain representation between (-1,1). This assumption is not feasible since there are infinite number of points inside this interval, which can not be supported by low memory devices with limited bit representation and can not be used for lossy media transmission. In~\cite{coutinho2018learning}, authors use a similar approach and reduces the model complexity and the input/output representation in order to force the model to achieve the XOR behavior between the data and the key. Although the results show that Alice and Bob can exchange confidential data successfully in less number of training iterations, however the model can be classified as semi linear, since it consists of only two fully connected layer with a single activation function. Thus, the model is imperfectly secured due to decreasing the encryption function. Moreover, the `mix and transform' architecture is absent due to the absence of convolution layers in the model, which makes the algorithm easy to be broken. In~\cite{fritschek2019deep}, the authors used autoencoder model and train it in a wiretap channel environment to achieve a key-less encryption between Alice and Bob. However, this approach is SNR dependant as Eve can decode the data if it has the same capability as Bob. The results showed that Eve did not reach to the uncertainty and was able to decode the confidential information partially as the SNR increases. 

\section{System Model}
\label{system_model}

The wiretap channel is an information-theoretic model for communication in the presence of an eavesdropper, which involves three nodes: sender (Alice), receiver (Bob) and eavesdropper (Eve) as shown in figure~\ref{fig:nn}. Alice encodes a confidential message $P$ using key, $K$, and outputs cipher data $C$, which is transformed to complex representation and transmitted as a complex vector $X$ to Bob. Both Bob and Eve  receive the complex cipher vector $Y$, which is $X$ after passing through the channel. The real representation $C'$ is recovered from $Y$. Bob decodes $C'$ using $K$ to obtain $\hat{P}_{Bob}$. However, Eve uses only $C'$ to obtain $\hat{P}_{Eve}$  which is Eve's predicted output for the confidential message $P$.
In this work, symmetric key encryption is considered where both Alice and Bob share the same key (K). Alice, Bob and Eve are all neural networks with parameters $\theta_{A}$, $\theta_{B}$ and $\theta_{E}$ respectively. In this paper, we investigate both clear and fading wiretap channels.

\subsection{Clear wiretap channel}
Clear wiretap channel is considered as a no loss transmission media. This type of encryption algorithm can be used to represent data for secure storage, and hence is applicable  to confidential data storage in untrusted third party cloud. 
In clear channel, if Alice transmits $X$, the received symbol for both Bob and Eve is $Y$, which can be given by:
\begin{equation}
    Y= X
    \label{Clear_model}
\end{equation}

\subsection{Gaussian wiretap channel}
Gaussian wiretap channel is defined as the channel with Additive White Gaussian Noise (AWGN) for both Bob and Eve. In other words, if Alice transmits $X$, the received symbol for both Bob and Eve, $Y$, is given by:
\begin{equation}
     Y= X + N \label{Guass_ch1}
\end{equation}
where $N\sim \mathcal{CN}(0,\,\sigma^{2})$ is the added complex noise vectors and $\sigma^{2}$ depends on the received signal to noise ratio (SNR).

\subsection{Rayleigh wiretap channel}
Similar to Gaussian wiretap channel, Rayeligh wiretap channel is defined as AWGN in nature, except the channel gain is not unity. If Alice transmits $X$, the received symbol for both Bob and Eve, $Y$, is given by:
\begin{equation}
     Y= HX + N \label{Ray_ch1}
\end{equation}
where $H$ is the channel matrix between the transmitter and the receiver.
In this work, we only consider flat fading channels, so $H$ is a diagonal matrix and the diagonal elements have a unity mean with Rayleigh distribution.

\subsection{Data and Key}

\textit{Range and Domain}:
Neural Network of Alice is designed to accept $P$ and $K$ in bits. In other words $ P,K \in \mathcal{B}$ where $\mathcal{B}=\{0,1\}$, while the cipher data $C \in \mathcal{R}$, where $\mathcal{R}$ is the set of real numbers. For practical implementation, we constrain 
$C$ within the range $(-1,1)$. Bob's network is designed to accept $C'$ as well as $K$ and outputs $\hat{P}_{Bob} \in \mathcal{R}$. Similar to Bob, Eve accepts $C'$ and outputs $\hat{P}_{Eve} \in \mathcal{R}$. We restrict $\hat{P}_{Bob}$ and $\hat{P}_{Eve}$ in the range between $(0,1)$. At the end of a successful training process, the values of $\hat{P}_{Bob}$ should converge to $\mathcal{B}$, while $\hat{P}_{Eve}$ should not. 
$\hat{P}_{Bob}$ and $\hat{P}_{Eve}$ are converted from $\mathcal{R}$ to $\mathcal{B}$ to extract the output bits. 

\textit{Length}:
The length of the data and key are essential parameters for data security. This is because the security introduced by the encryption algorithm depends on the length of the data as well as the key. According to Shannon secrecy~\cite{shannon1949communication}, the system can be perfectly secure if the key size equals to the data size such that:
\begin{equation}
\lim_{N_{K}\to N_{P}} I(C,P)=0    
\end{equation}
where $I(C,P)$ is the mutual information between cipher data and plain data, and $N_{P}$ and $N_{K}$ are the lengths of the plain data and the key respectively. Thus, one-time pad encryption~\cite{one_time} is information-theoretic secure because the lengths of the data and the key are equal. In this work, we have trained the network on a finite set of keys (i.e $N_{K} < N_{P}$)  such that the networks are not restricted to learn only one-time pad encryption.
The scope of this paper is limited to single carrier communication but we have designed the system parameters in a way that it can be easily adapted to OFDM systems. Hence, the block sizes of $K$, $P$ and $C$ equal to $N$, where $N$ is double the size of the FFT. The search space of the cipher symbol $S(C)$ can be given by:
\begin{equation}
    S(C)= 2^{N}
    \label{eq:search space}
\end{equation}
This indicates that larger the size of the FFT, the search space that it might get mapped to increases, which increases the secrecy capacity for $C$. This property is enhanced as we move from bit-level modifications in traditional higher layer security to real number domain with $C \in \mathcal{R}$.


\section{Problem statement}
\label{formulate}
In this section, we define the objective of each member of the network. Alice and Bob try to exchange confidential information in a secure way such that Eve can not recover plain data $P$ 
from the cipher data $C'$. On the other side, Eve tries to reconstruct ${P}$ from $C'$, which can be achieved by reducing the error between $\hat{P}_{Eve}$ and $P$. Informally, the objective of Alice and Bob is to figure out a secure way to exchange the confidential data as well as defeat Eve to recover any information from the shared cipher text. Based on these objectives, the three neural networks should be trained in an adversarial manner. The loss functions for both Alice and Bob, as well as Eve, have to be derived to support the adversarial behavior of Eve.

We define $A(\theta_{A},P,K)$, $B(\theta_{B},C,K)$ and $E(\theta_{E},C)$ as the output vectors of Alice, Bob and Eve respectively. In addition $d(P,\hat{P}) = \sqrt{\sum_{i=1}^{N}(P_{i}-\hat{P_{i}})^2}$ (i.e $d(P,\hat{P})$ is the L2 norm in case of vectors and Frobenius Norm in case of matrices).

Intuitively, the loss function of Eve is derived as:
\begin{multline}
    \label{loss_eve}
    L_{E}(\theta_{A},\theta_{E},P,K)=E_{P,K}(d(P,\hat{P}_{Eve})=\\ E_{P,K}(d(P,E(\theta_{E},A(\theta_{A},K,P))))
\end{multline}
It is to be noted here that we train Eve on the same plain data $P$ as Alice and Bob to minimize the loss function.
Similarly, the loss function of Bob is derived as:
 \begin{multline}
  \label{loss_Bob}
    L_{B}(\theta_{A},\theta_{B},P,K)=
    E_{P,K}(d(P,\hat{P}_{Bob}))\\=E_{P,K}(d(B(A(\theta_{A},K,P),K,\theta_{B})),P) 
 \end{multline}
From \ref{loss_Bob}, it is inferred that Bob's Loss function depends on the cipher data and the shared key as well; however, this is not sufficient to train the network in an adversarial manner. Thus, the adversarial Loss function of Bob should minimize the error between $\hat{P}_{Bob}$ and $P$ as well as maximize the error between $\hat{P}_{Eve}$ and $P$. This problem is similar to min-max optimization in GAN~\cite{AE_GANS}. In order to derive this, we have to optimize a joint loss function between Alice and Bob to update their parameters simultaneously and is given by:
 \begin{multline}
    \label{loss_total}
    L_{A,B}(\theta_{A},\theta_{B})=argmin_{\theta_{A},\theta_{B}}(L_{B}(\theta_{A},\theta_{B},P,K)\\-L_{E}(\theta_{A},\theta_{E},P,K)) 
 \end{multline}

From (\ref{loss_eve}) and (\ref{loss_total}), it is inferred that both of them depend on $\theta_{A}$. However, cooperative learning happens only between Alice and Bob to defeat Eve. So during the training phase, $\theta_{A}$ only gets updated jointly with Bob, and they are frozen during the training of Eve in each epoch.
According to the definition of entropy~\cite{shannon1948mathematical}, the receiver reaches the maximum uncertainty, if the received value of Eve equals to $0.5$. Once that is reached, random guessing is the only way for Eve to recover the transmitted bits. Accordingly, the uncertainty property has to be added to equation (\ref{loss_total}), such that both Alice and Bob can try to figure out a transmission pattern that satisfies the maximum uncertainty to Eve. Thus equation(\ref{loss_total}) can be reformulated as:
\begin{multline}
    \label{loss_total_mod}
    L_{A,B}(\theta_{A},\theta_{B})=argmin_{\theta_{A},\theta_{B}}(L_{B}(\theta_{A},\theta_{B},P,K)\\+(0.5-L_{E_{N}}(\theta_{A},\theta_{E},P,K))^2)
 \end{multline}
where $L_{E_{N}}$ is the normalized loss function of Eve. 
As shown in (\ref{loss_total_mod}), the first component tends to minimize the error between Alice and Bob, while the later one enforces the mean loss of Eve to be $0.5$. Thus the received values take the value of $0.5$, which increases the uncertainty at  Eve's side. Recall that we use hard decision decoding to convert the received data values to data bits.


\section{Neural Network Design}
\label{NN_structure}

\subsection{Neural network structure}

Figure~\ref{fig:nn} shows the neural network architecture used by the three entities in the system. Alice accepts $P$ and $K$ representing plaintext and key respectively. Bob accepts $K$ and $C'$, which is the cipher text signal after passing through the channel. Eve's input is only $C'$. The output of Alice is the cipher data $C$, whereas the output of Bob and Eve are $\hat{P}_{Bob}$ and 
$\hat{P}_{Eve}$  denoting the predicted $P$ for Bob and Eve respectively.
All the input and output parameters, $P$, $K$, $C$, $C'$, $\hat{P}_{Bob}$ and $\hat{P}_{Eve}$ are vectors of dimension $N$.
We utilize a `mix and transform' architecture to build the three neural networks. The network starts with fully connected layers (FC) without any activation function being introduced. The purpose of this layer is mixing the key and the data bits so that the output bits are permuted input bits or a mix between data and key bits. The network consists of multiple convolutional layers to enable squeezing data and key bits. 
The convolutional layer is described as $conv(W, d_{in}, d_{out}, s)$, where $W$ is the window size, $d_{in}$ is the input depth, $d_{out}$ is the output depth and $s$ is the stride. The stride is defined as the number of steps the window is shifted. In general, the convolutional layers are used to extract the features in image classification by neural networks. Thus, in cryptography applications, convolutional layers are used to extract the common features between the data and the key. For Alice, sigmoid activation function is used after each convolutional layer which is given by:
\begin{equation}
    \sigma(z)= \frac{1}{1+e^{-z}}
    \label{sig}
\end{equation}
while the final layer, $tanh$ activation function is used to make the transmitted data a bipolar form. $tanh$ output ranges between (-1,1) and is given by:
 \begin{equation}
     tanh(x)=\frac{e^{x}-e^{-x}}{e^{x}+e^{-x}}
     \label{tanh}
 \end{equation}
On the other hand, we use Relu activation function \cite{DL_book} at the first layers for Bob and Eve to compensate the channel effect in the forward path and increase the learning propagation to Alice layers. Sigmoid activation function has been used so that the output converges between (0,1) to make the output vectors achieve the bit values at the end of the training process. 
A hard decision decoding is performed to transform $\hat{P}_{Bob}$ and $\hat{P}_{Eve}$ from $\mathcal{R}$ to $\mathcal{B}$.

\subsection{Discrete activation function} 
Most of the neural network (NN) training algorithms are based on gradient descent methods. This has moved all the research related to the NN to use floating point number representation and smooth activation functions to guarantee the existing gradient values to the NN parameters to achieve the cost function convergence during the training \cite{Dis}. This is because the updated NN weights during training is given by:
\begin{equation}
    w_{i}^{k+1}= w_{i}^{k}-\epsilon\frac{\partial J}{\partial w_{i}}
    \label{weights}
\end{equation}
where $w_{i}$ is the NN weight, $\epsilon$ is the learning rate, $k$ is the training epoch and $\frac{\partial J}{\partial w_{i}}$ is the partial derivative of the cost function $J$ with respect to the NN weight $w_{i}$. 

\begin{figure}
    \centering
    \begin{subfigure}[b]{0.24\textwidth}
        \includegraphics[width=\textwidth]{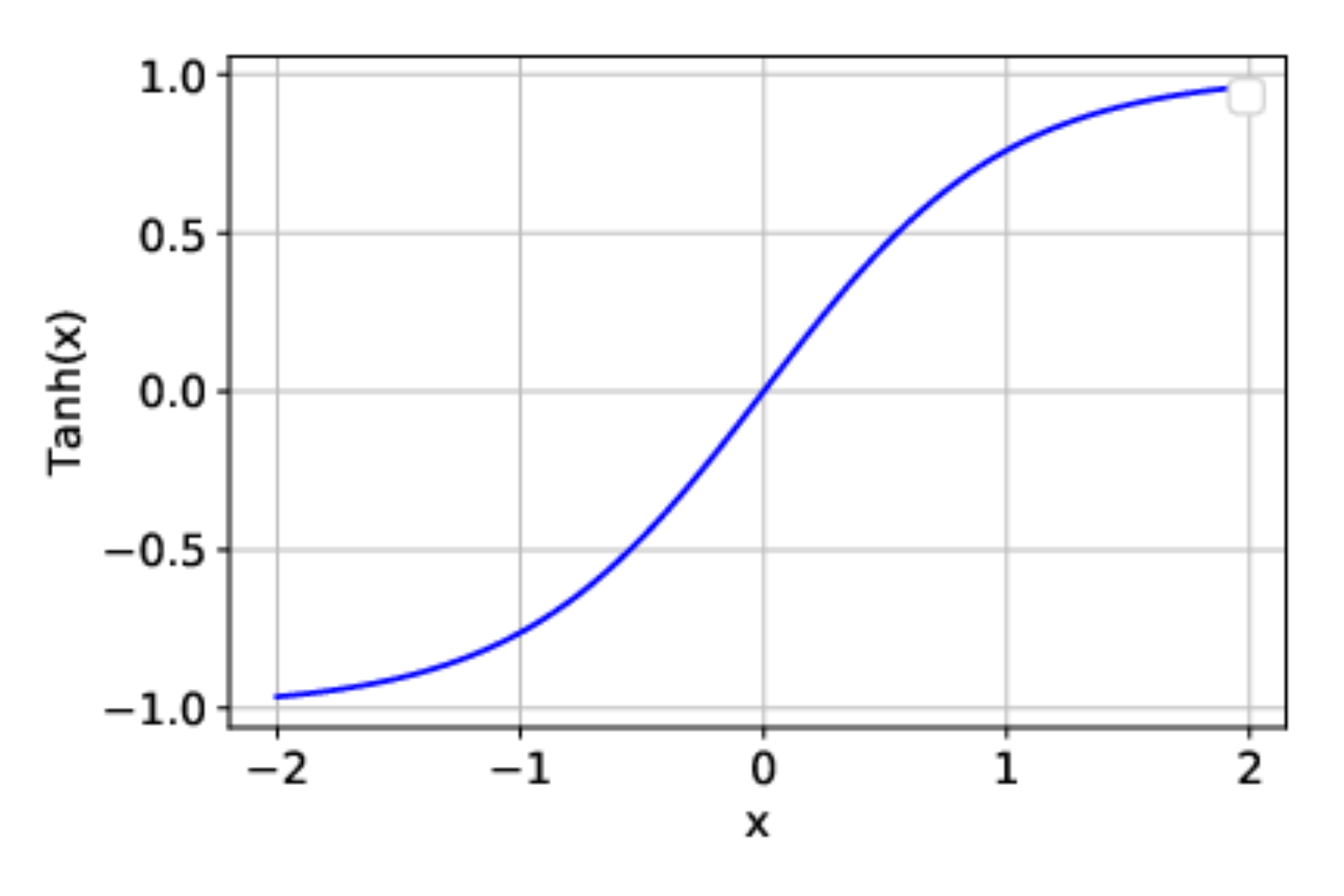}
        \caption{Continuous $tanh$.}
        \label{fig:tanj}
    \end{subfigure}
    \begin{subfigure}[b]{0.24\textwidth}
        \includegraphics[width=\textwidth]{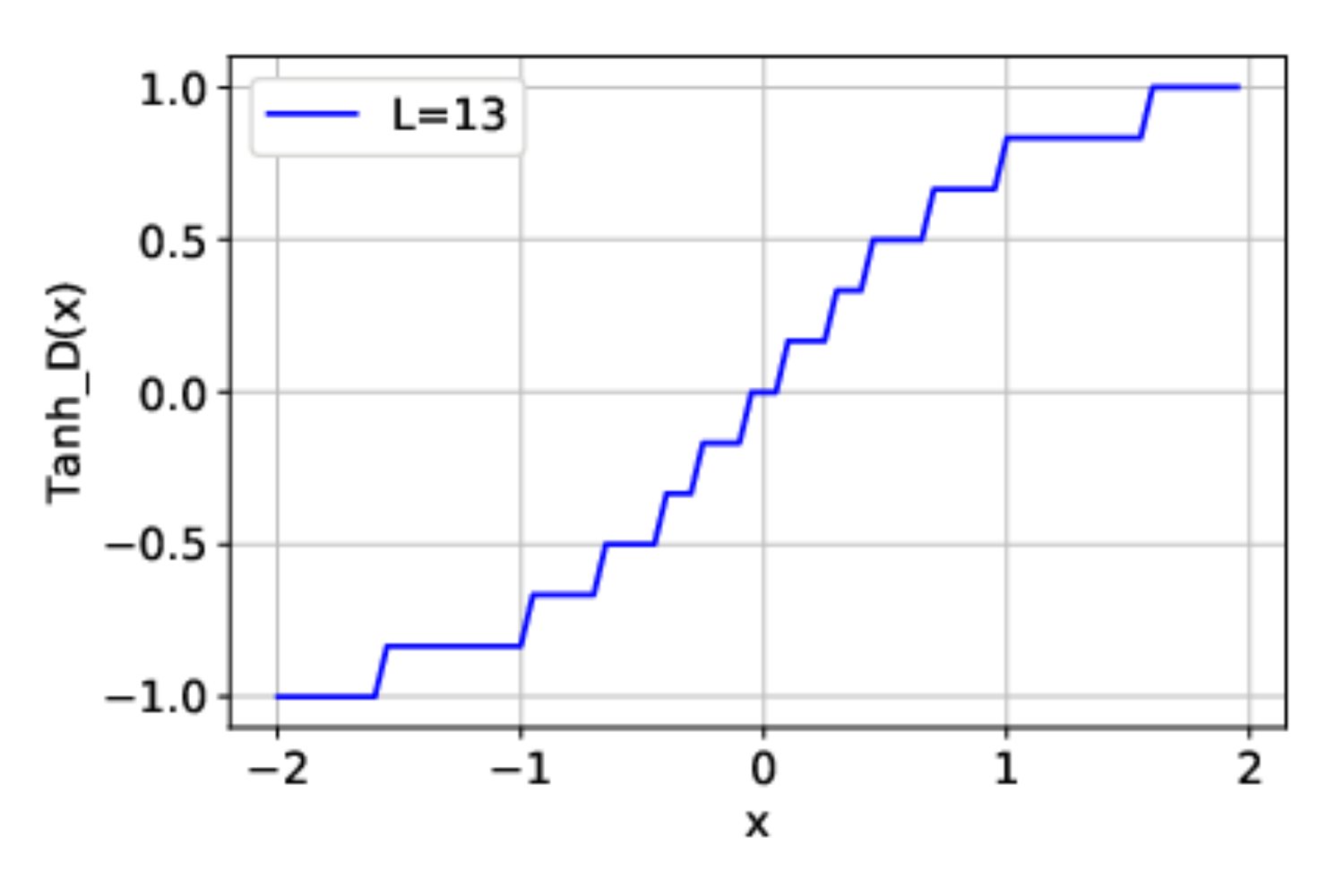}
        \caption{Discrete $tanh$ with 13 levels.}
        \label{fig:discrete tanh}
    \end{subfigure}
    \caption{Continuous and discrete $tanh$ activation functions.}
    \label{fig:Activation_function}
\end{figure}

As shown in (\ref{Guass_ch1}) and (\ref{Ray_ch1}), the added corruption in noisy lossy channel is considered as a continuous random variable. Hence, $X$ has to be at a discrete level to eliminate the added corruption at the receiver. In other words, the receiver will not be able to compensate the channel effect if the minimum distance between the transmitted constellation points is too small.
In addition, floating point representation is impractical for limited memory applications. 
On the other hand, adding discrete layers in the neural networks causes error in back propagation, since the gradient of any discrete function is undefined. Consequently, the total gradient of the network will vanish, and the cost function will not converge. In this work, we define a discrete $tanh$ function (i.e. $y=tanh_D(x)$) to be used to quantify the output of Alice NN into defined levels as shown in Algorithm (\ref{TanhD}). In order to use $tanh_D(x)$ with the regular gradient methods, we define a suitable derivative to be used in the backward path, which is given by:
\begin{equation}
    \frac{\partial (tanh_{D}(x))}{\partial x}= 1.0- \tanh^{2}{(x)}
\end{equation}
In this work, we do not perform the quantization on the input of the activation function. Thus, the output of the last convolutional layer can take any value in the $tanh_{D}(x)$ domain (i.e. the firing region of the activation function), which enables the gradients to propagate to the earlier layers during the training process. Figure~\ref{fig:Activation_function} shows the continuous and discrete $tanh$ functions used for clear and Gaussian wiretap channels respectively. As the number of levels increases, the behavior of $tanh_D(x)$ tends to be similar to $tanh(x)$. Thus it is not feasible to use $tanh_D(x)$ with a higher number of levels in transmission over lossy media. On the other hand, if the number of levels decreases, the back propagation error increases, which leads to exploding gradient problem~\cite{vanish}. In this work, we trained the system on various values of the number of levels $L$ to find the optimal number of levels $L^{*}$.

\vspace*{-5pt}
\begin{algorithm}
\SetAlgoLined
\KwResult{ $y_q$}
 $y= tanh(x)$\;
 $y_{min}= -1$\;
 $y_{max}= 1$\;
 $step= (y_{max}-y_{min})/(levels-1)$\;
 $y_{q}=\floor*{(y-y_{min})/step+0.5}*step+y_{min} $
 \caption{$y_{q}=tanh_D(x,levels)$.}
 \label{TanhD}
\end{algorithm}

\begin{figure}
    \centering
    \begin{subfigure}[b]{0.23\textwidth}
        \includegraphics[width=\textwidth]{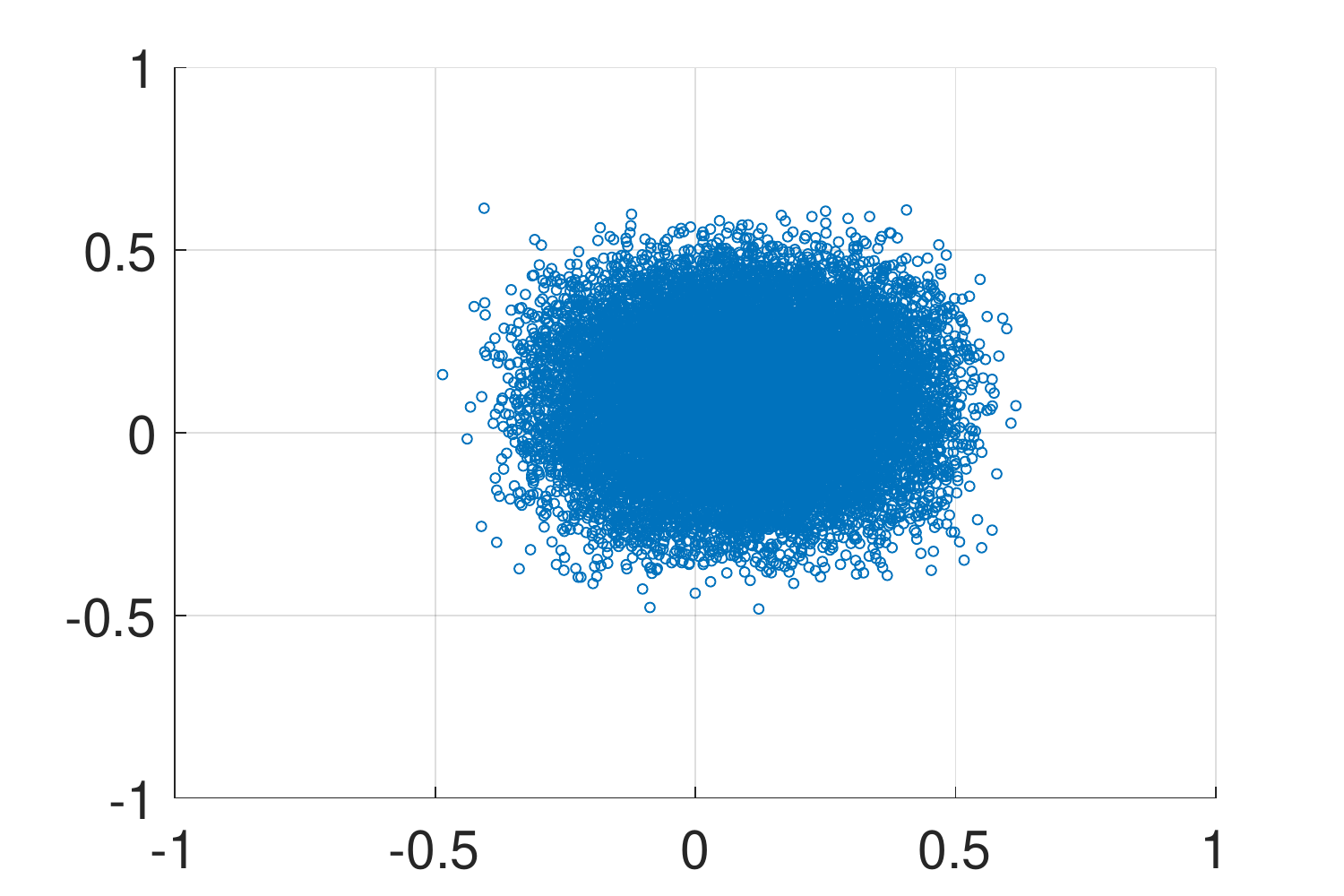}
        \caption{$tanh$ function followed by modulation in clear wiretap channel.\\}
        \label{fig:modulate_tanh}
    \end{subfigure}
    \quad
    \begin{subfigure}[b]{0.23\textwidth}
        \includegraphics[width=\textwidth]{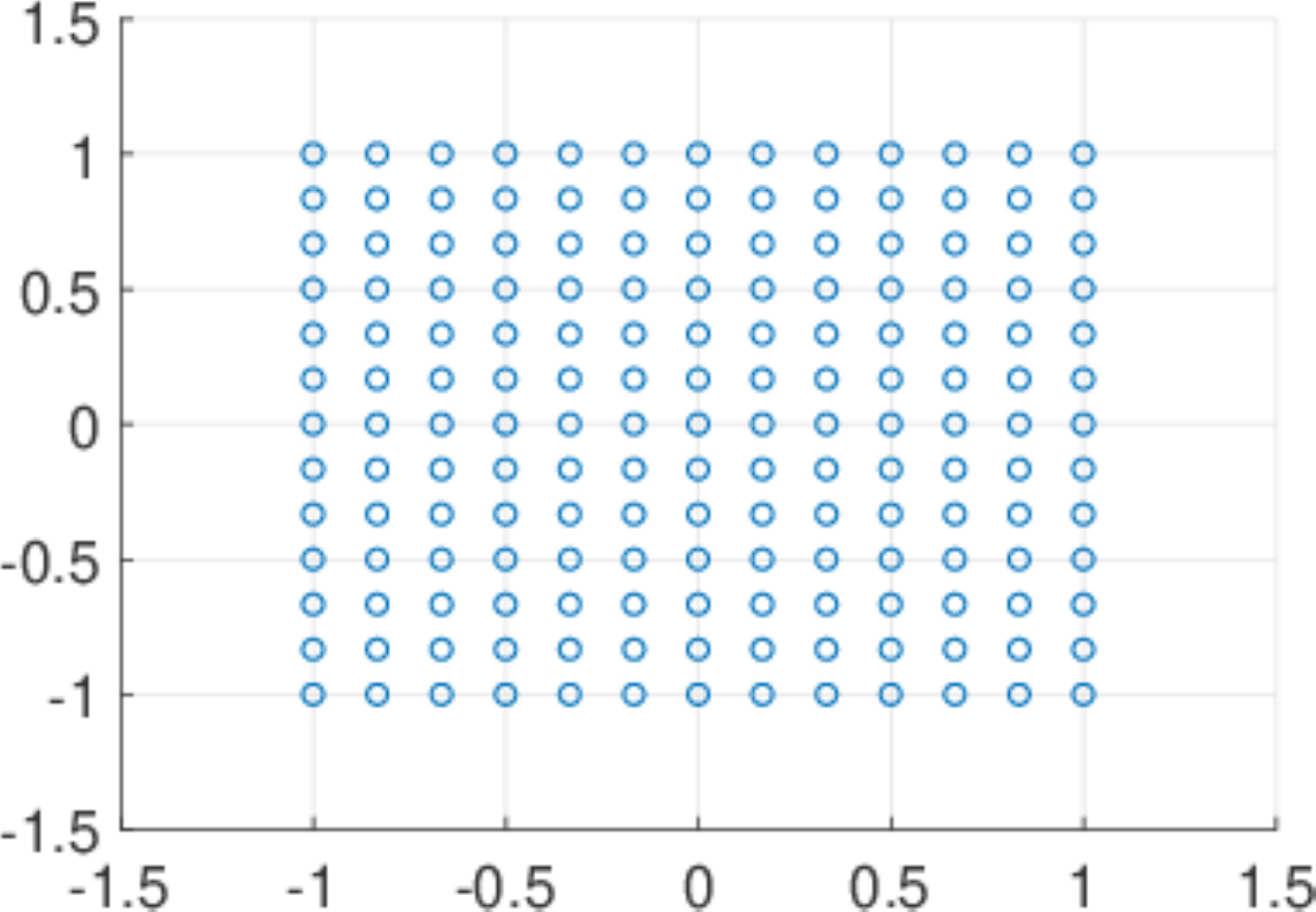}
        \caption{Discrete $tanh$ function followed by modulation in Gaussian and Rayleigh wiretap channels.}
        \label{fig:modulate_d_tanh}
    \end{subfigure}
    \caption{Cipher data constellation $X$.}
    \label{fig:constellation}
\end{figure}

\subsection{Modulate and Demodulate}

The NN supports real numbers only; however this representation is not efficient for carrier transmission as we are not utilizing the real and imaginary domains. In this work, we convert the real representation of the transmitted symbol $C$ to complex transmitted symbol $X$ to support single or multi-carrier transmission. Each complex sample $x$ in the complex vector $X$ can be given by:
\begin{equation}
    x_{m}= c_{i}+jc_{i+1} 
\end{equation}
where $c_i$ is the $i^{th}$ real sample in $C$, $m$ takes values form 1 to $\ceil{\frac{N}{2}}$ 
 and $i$ ranges from 1 to $N$.
At the receiver side (i.e. either Bob or Eve), the reverse operation takes place, such that $Y$ is changed to $C'$ at the NN input.
The result of modulation is represented by the cipher data constellation, as shown in figure~\ref{fig:constellation} with both $tanh$ and $tanh_D$ functions. With clear channel, the signal can take any real value, which shows the Gaussian distribution of the constellation. In the presence of noise, as we introduce discrete steps, the resultant constellation takes the form of QAM signal.  


\subsection{Parameter initialization} 
In this work, we use Xavier initialization~\cite{Xavier_initialization} to initialize the total weights and biases. This initialization is used to accelerate the convergence of the neural networks and avoid gradient saturation. Moreover, we did not restrict the network to learn a defined function such as XOR; however, we leave the network to learn the secured modulation waveform such that it minimizes the cooperative learning loss function.

 \begin{figure*}
    \centering
    \begin{subfigure}[b]{0.24\textwidth}
        \includegraphics[width=\textwidth, height= 2.7 cm]{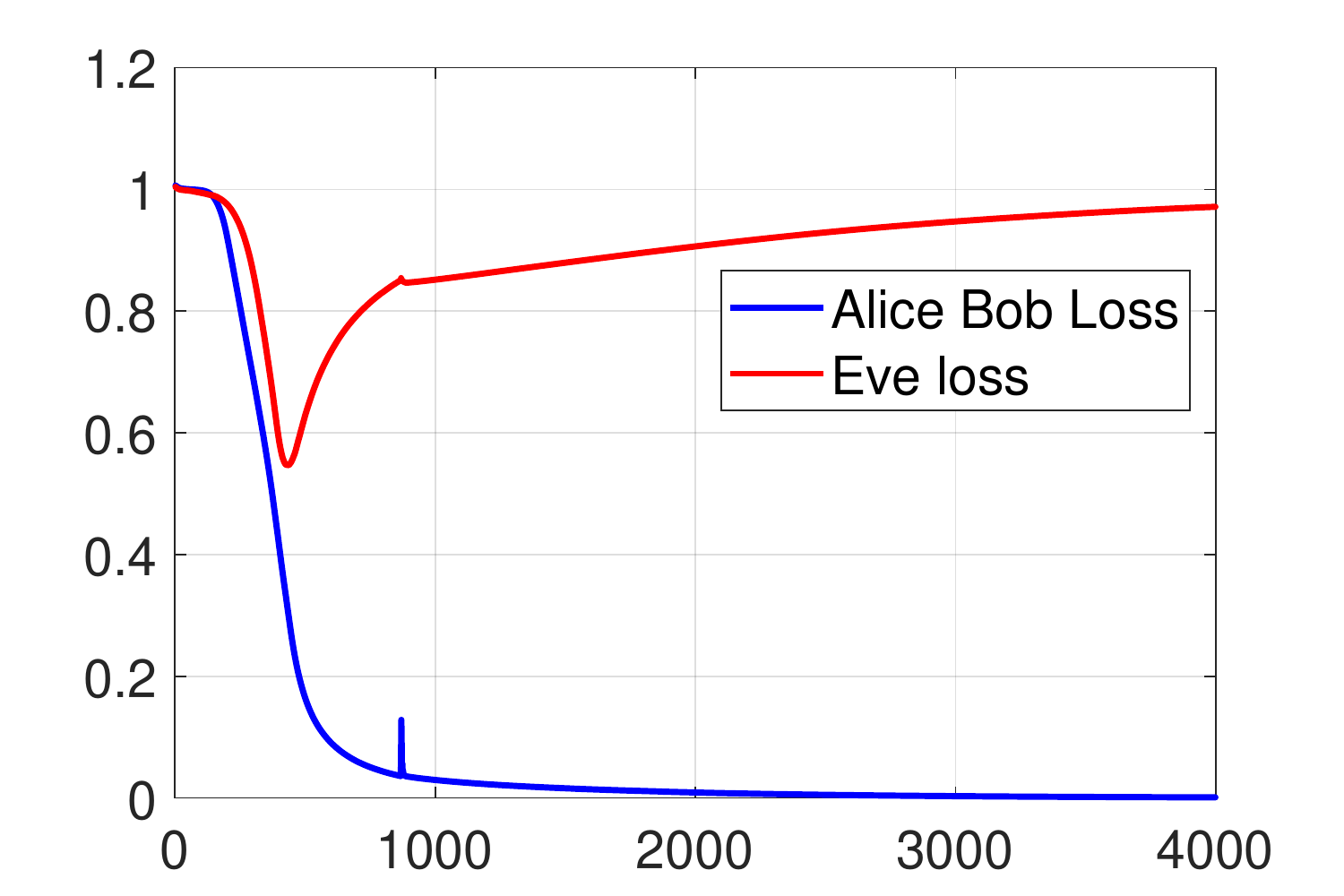}
        \caption{Loss function.}
        \label{fig:loss_clear}
    \end{subfigure}
    \begin{subfigure}[b]{0.24\textwidth}
        \includegraphics[width=\textwidth]{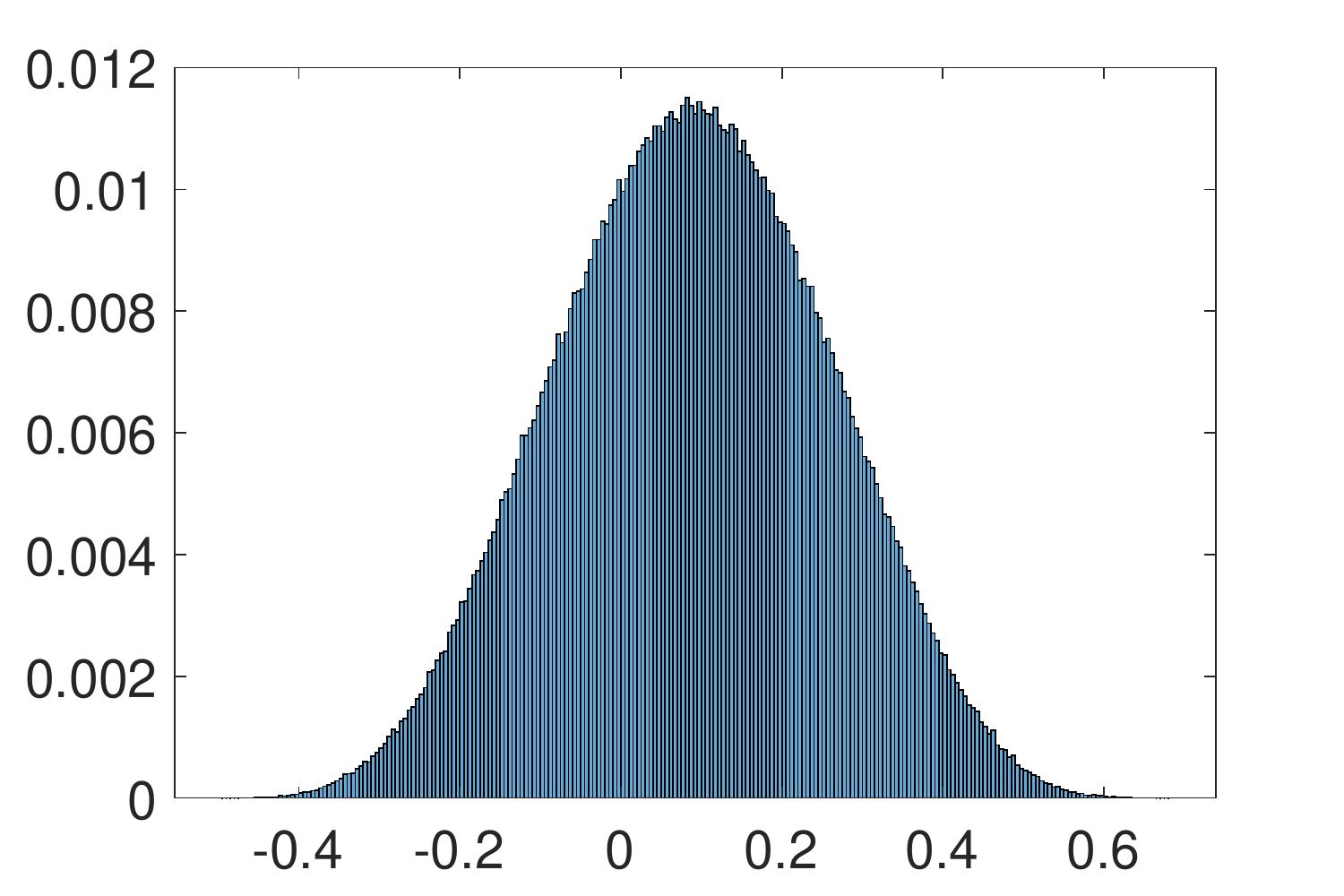}
        \caption{$C$ distribution.}
        \label{fig:cipher_distr_clear}
    \end{subfigure}    
    \begin{subfigure}[b]{0.24\textwidth}
        \includegraphics[width=\textwidth]{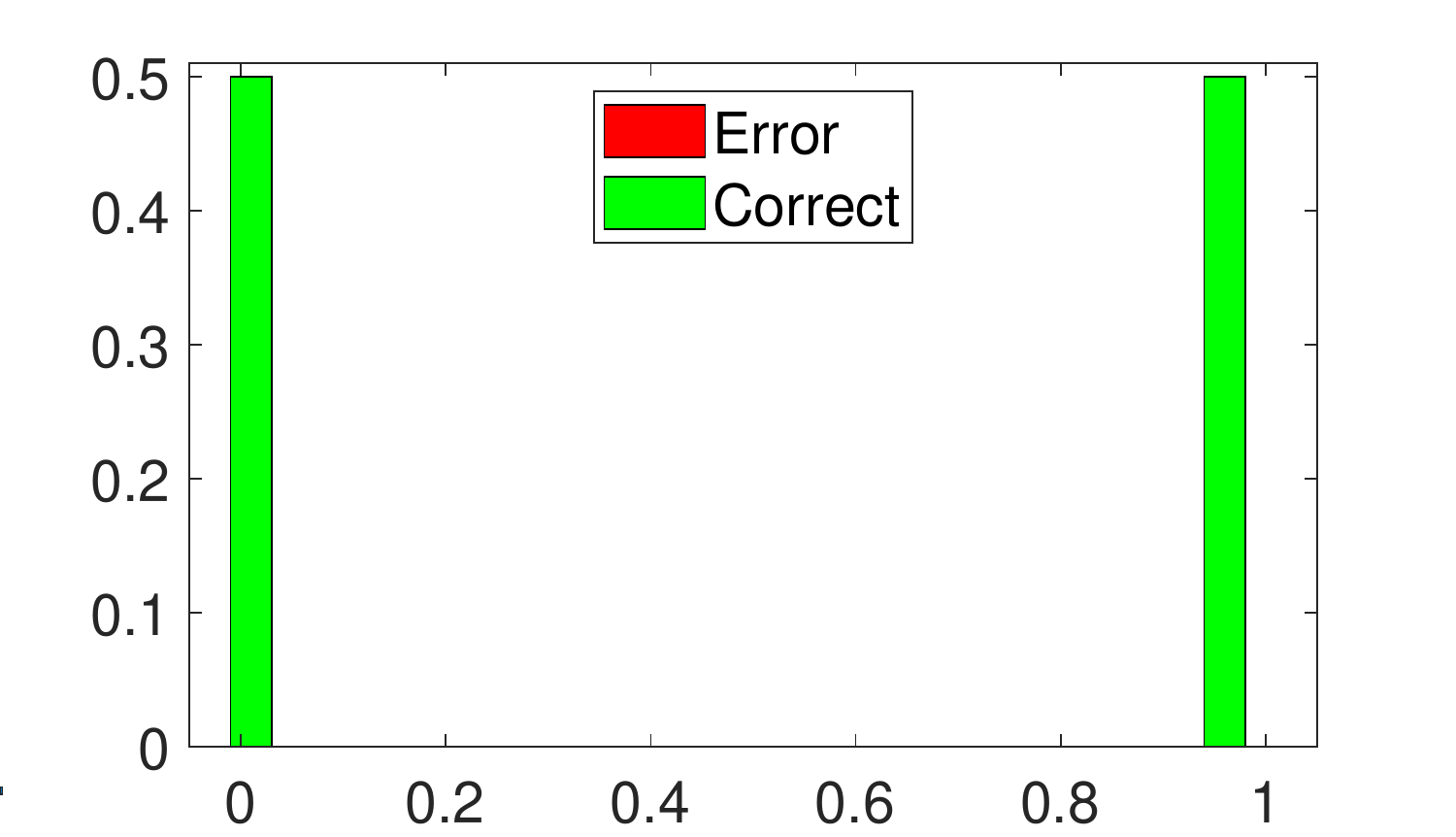}
        \caption{$\hat{P}_{Bob}$ distribution.}
        \label{fig:Bob_distr_clear}
    \end{subfigure}
    \begin{subfigure}[b]{0.24\textwidth}
        \includegraphics[width=\textwidth]{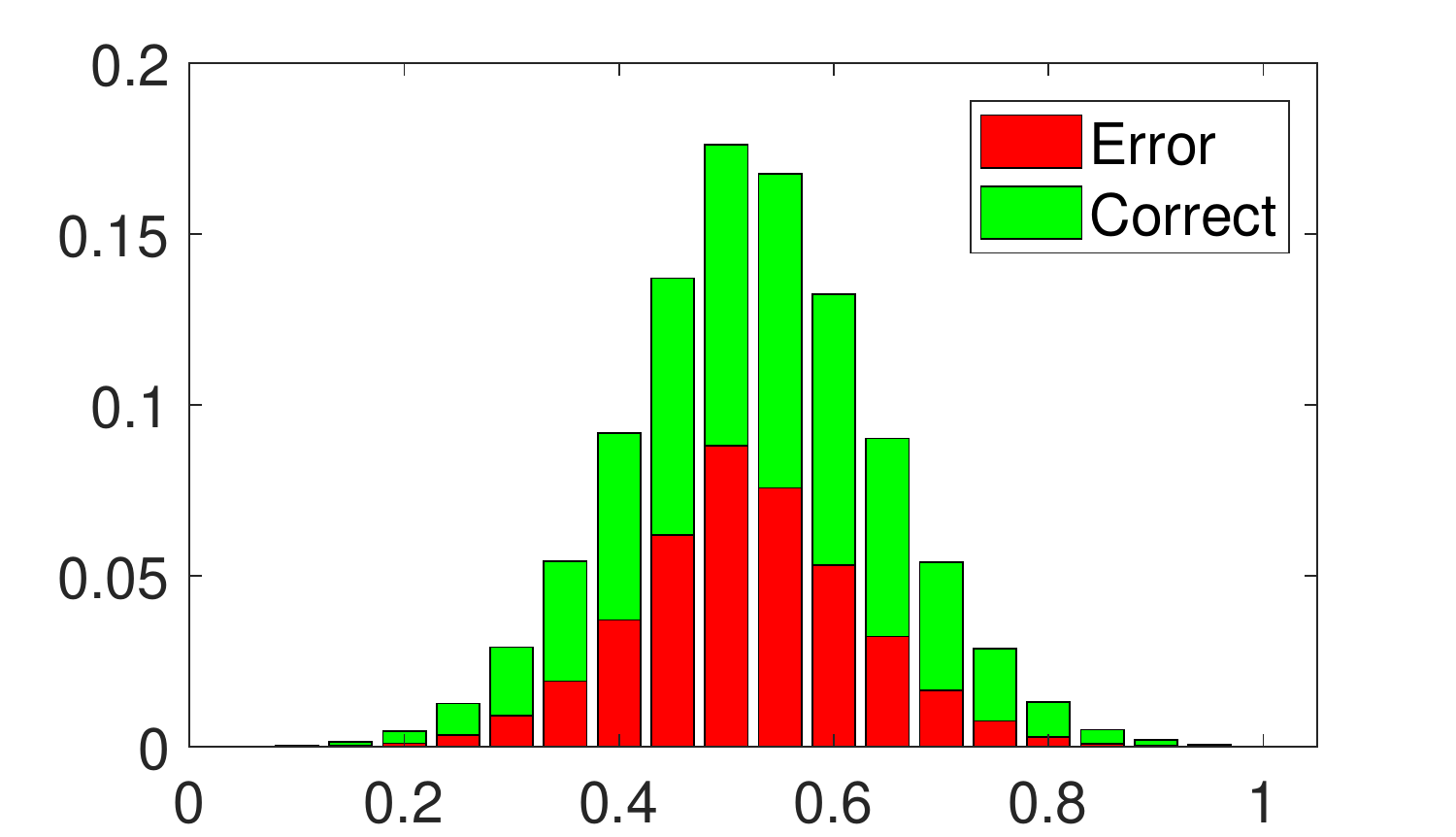}
        \caption{$\hat{P}_{Eve}$ distribution.}
        \label{fig:Eve_distr_clear}
    \end{subfigure}
    \caption{Loss function and data distribution for different entities with $tanh$ function in the last layer.}
    \label{fig:Loss_Dist_clear}
\end{figure*}

\begin{figure*}
    \centering
    \begin{subfigure}[b]{0.24\textwidth}
\includegraphics[width=\textwidth,height= 2.7cm]{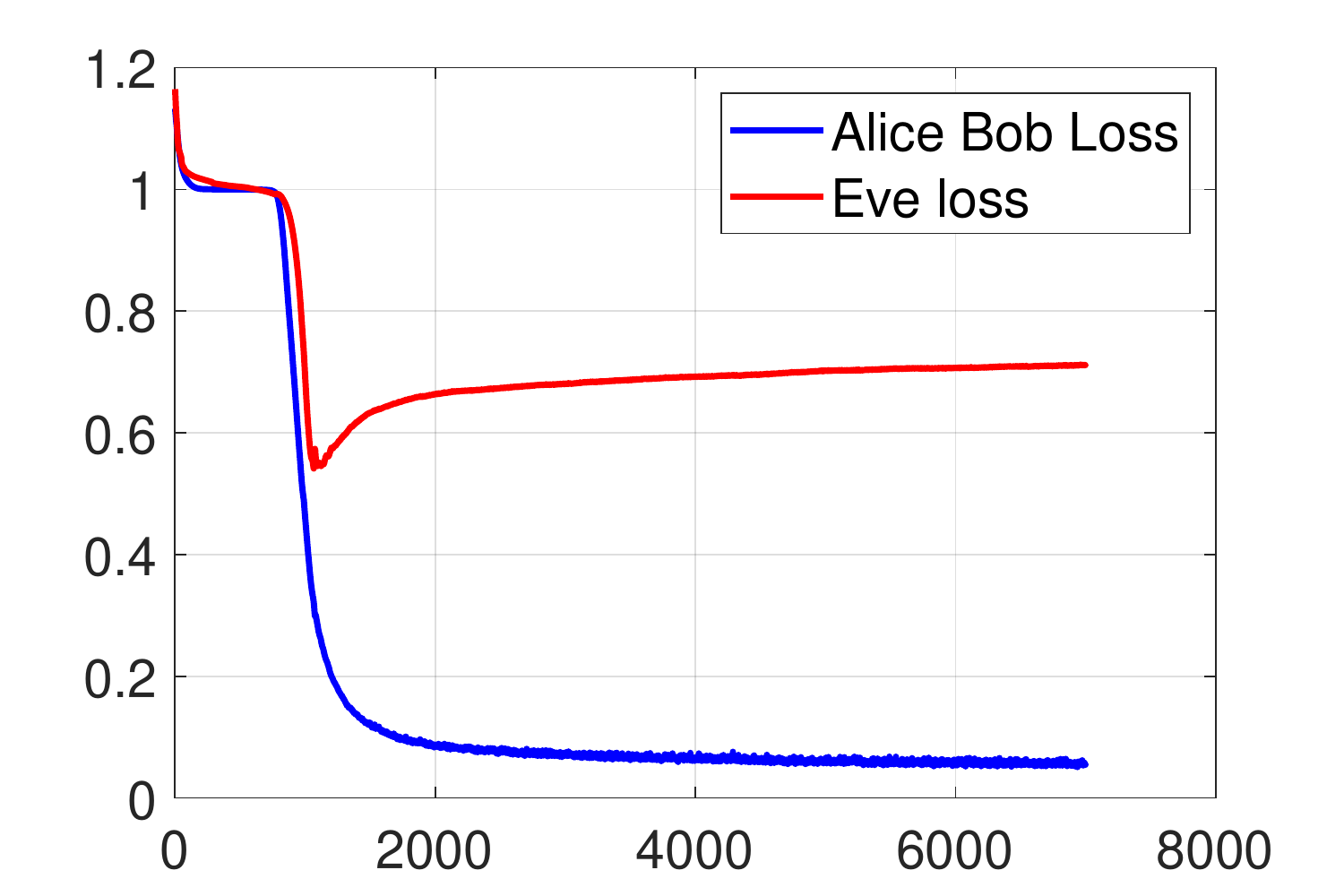}
        \caption{Loss function.}
        \label{fig:Loss_Q}
    \end{subfigure}
    \begin{subfigure}[b]{0.24\textwidth}
        \includegraphics[width=\textwidth]{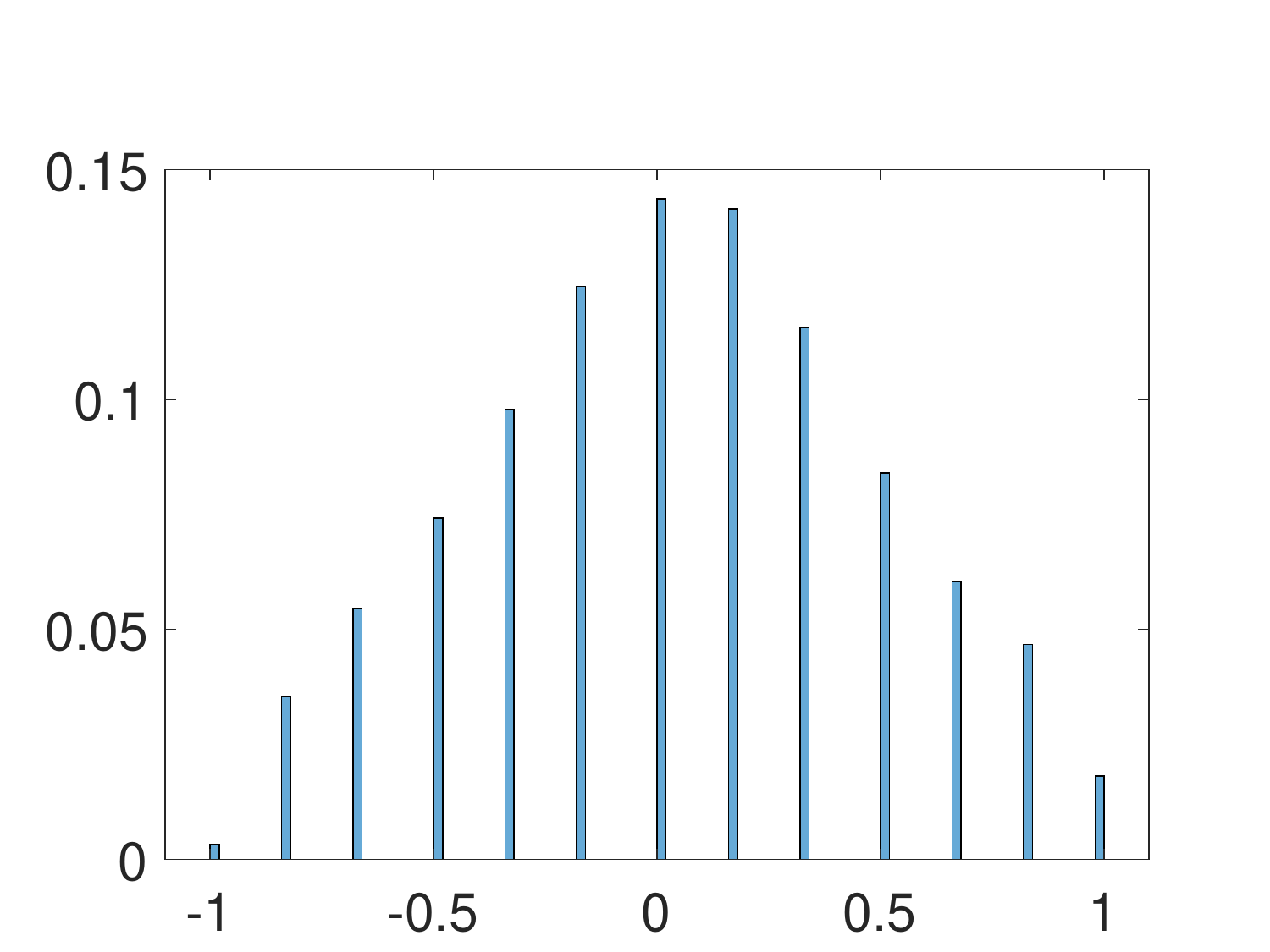}
        \caption{$C$ distribution.}
        \label{fig:cipher_distr}
    \end{subfigure}
    \begin{subfigure}[b]{0.24\textwidth}
        \includegraphics[width=\textwidth]{Figures/Bob_dis_awgn2.pdf}
        \caption{$\hat{P}_{Bob}$ distribution.}
        \label{fig:Bob_distr}
    \end{subfigure}
    \begin{subfigure}[b]{0.24\textwidth}
        \includegraphics[width=\textwidth]{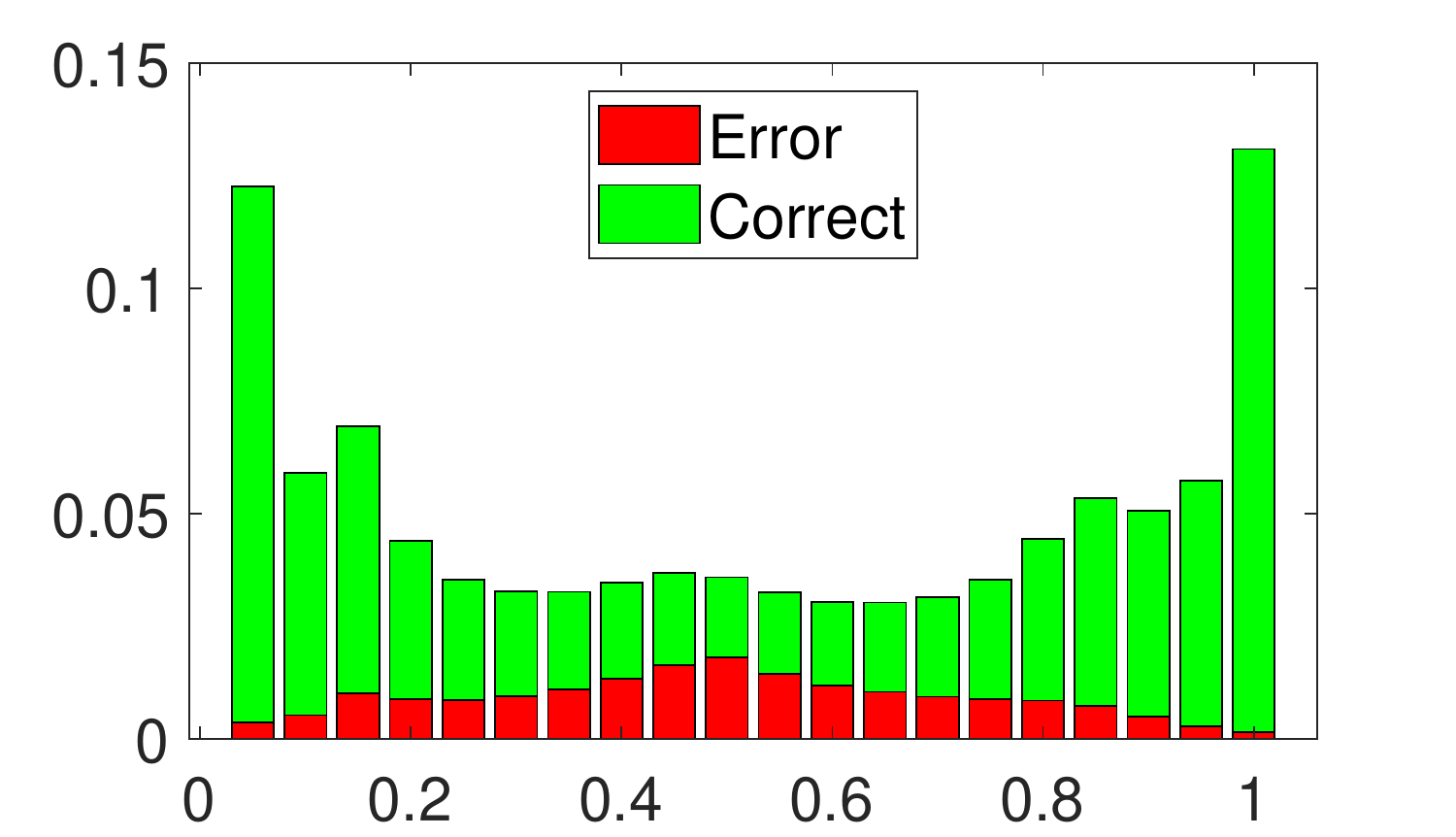}
        \caption{$\hat{P}_{Eve}$ distribution.}
        \label{fig:Eve_distr}
    \end{subfigure}
    \caption{Loss function and data distribution for different nodes with Discrete $tanh$ function in Gaussian wiretap channel.}
    \label{fig:Loss_Q_distr}
\end{figure*}

\begin{figure*}
    \centering
    \begin{subfigure}[b]{0.24\textwidth}
        \includegraphics[width=\textwidth, height=2.7 cm]{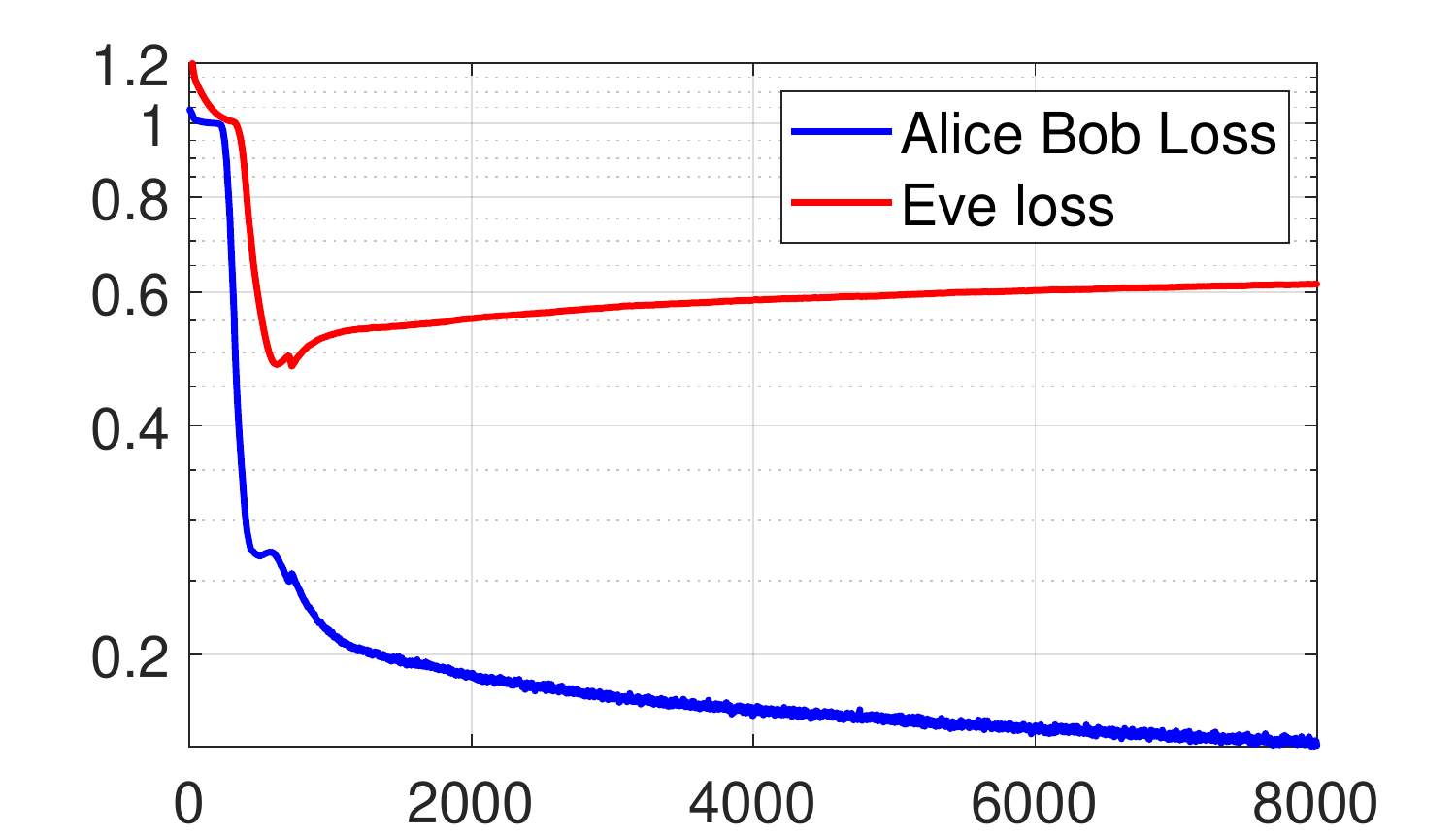}
        \caption{Loss function.}
        \label{fig:Loss_Q_ray}
    \end{subfigure}
    \begin{subfigure}[b]{0.24\textwidth}
        \includegraphics[width=\textwidth]{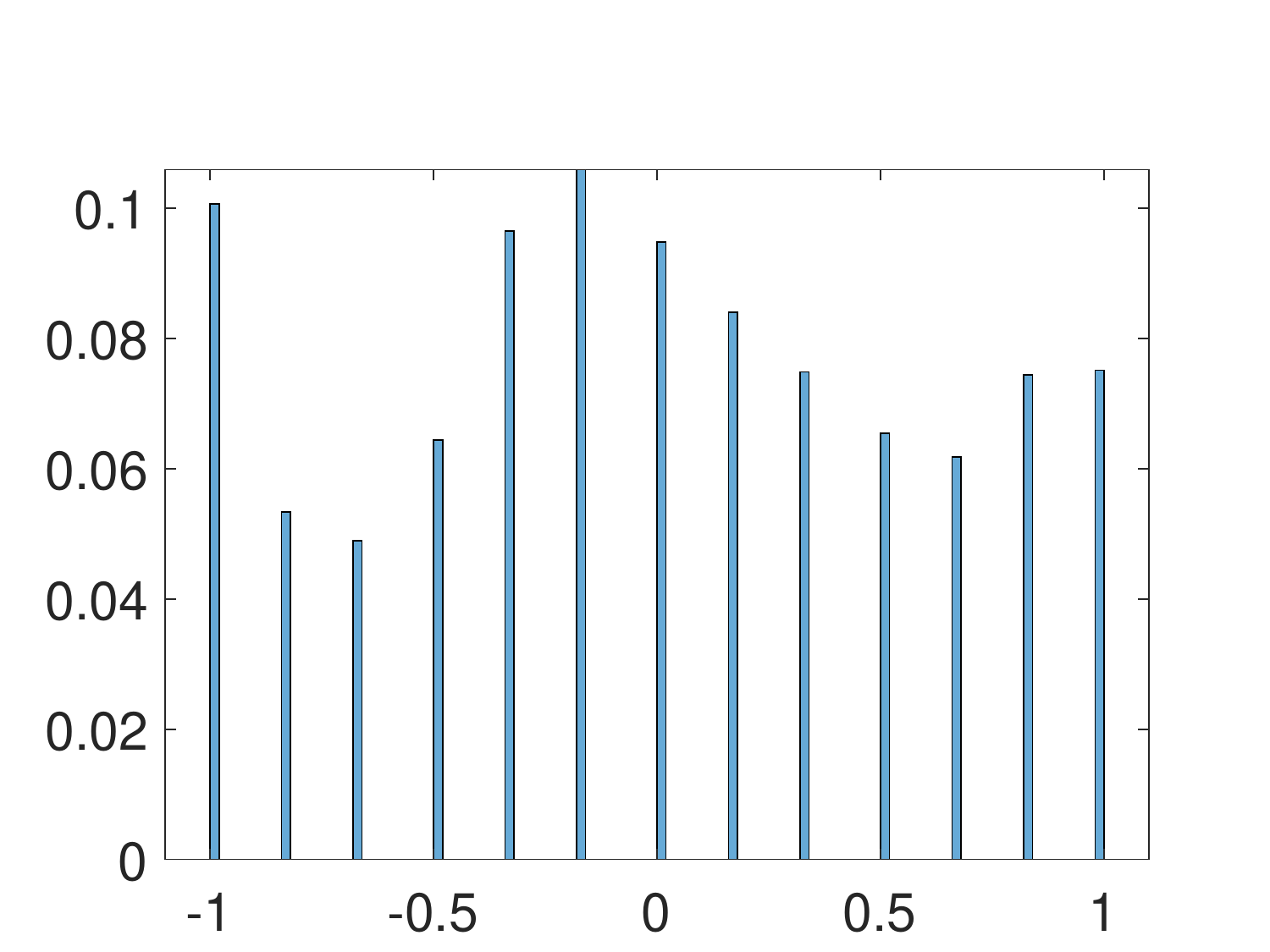}
        \caption{$C$ distribution.}
        \label{fig:cipher_distr_ray}
    \end{subfigure}
    \begin{subfigure}[b]{0.24\textwidth}
        \includegraphics[width=\textwidth]{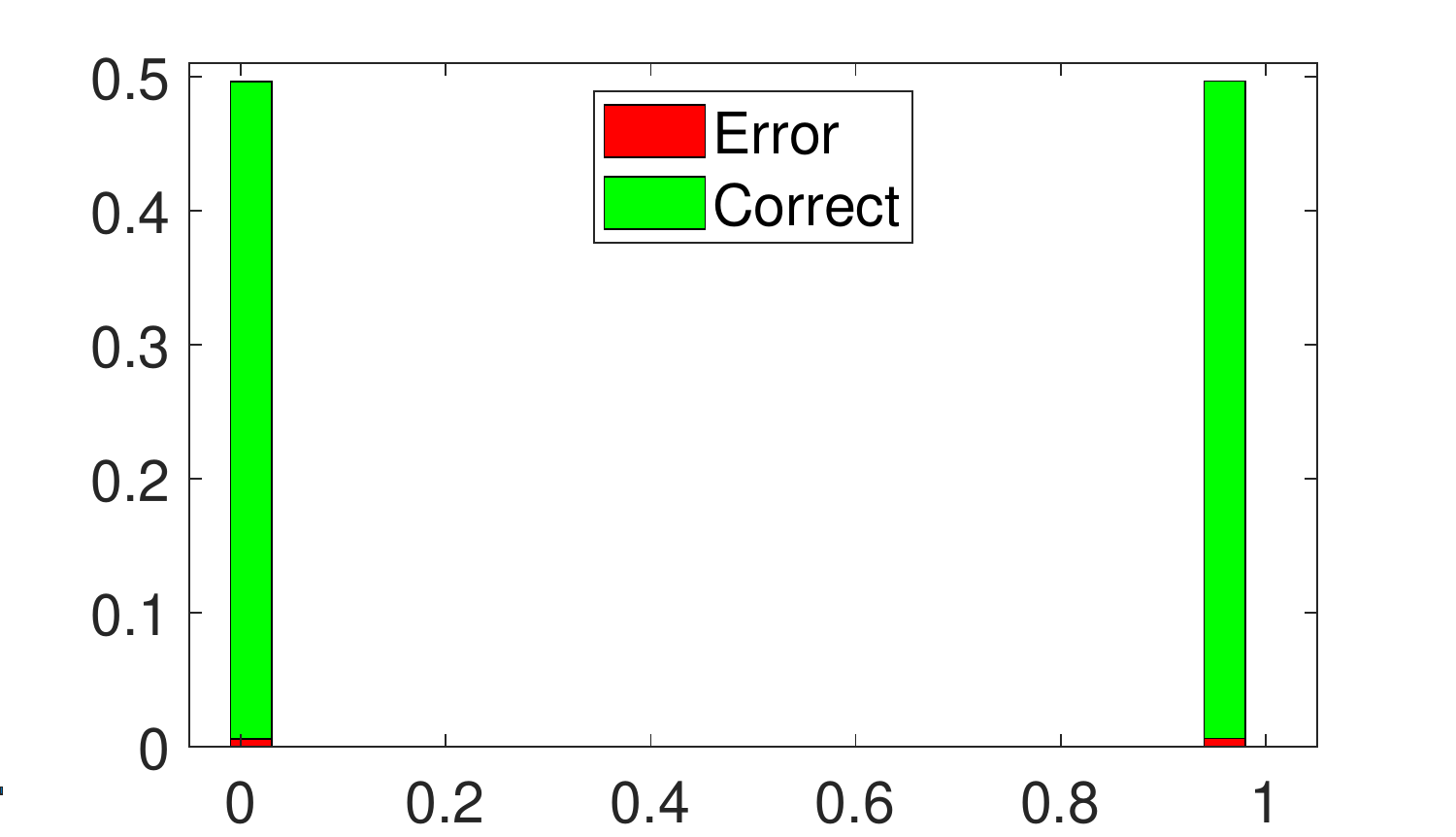}
        \caption{$\hat{P}_{Bob}$ distribution.}
        \label{fig:Bob_distr_ray}
    \end{subfigure}
    \begin{subfigure}[b]{0.24\textwidth}
        \includegraphics[width=\textwidth]{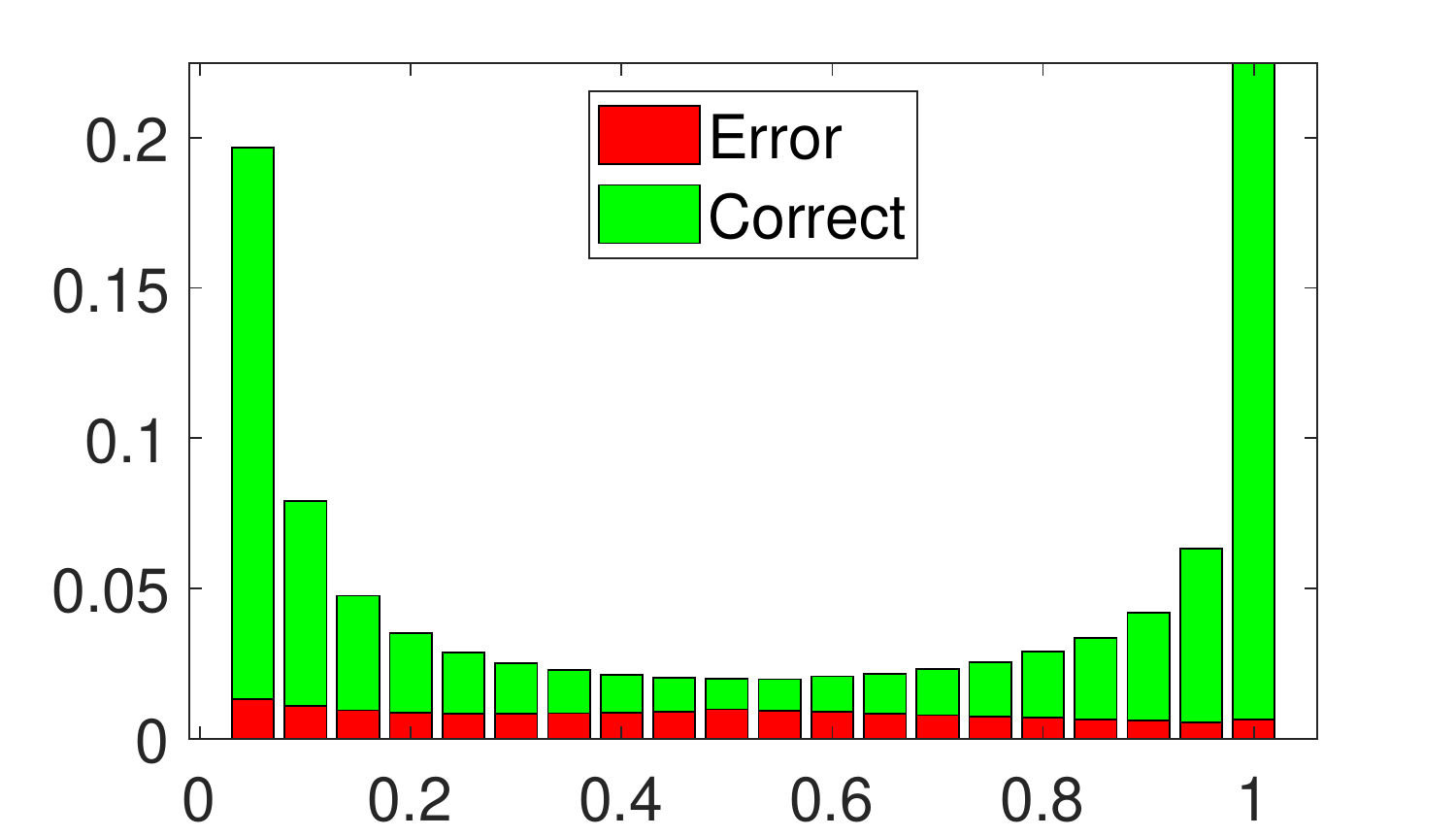}
        \caption{$\hat{P}_{Eve}$ distribution.}
        \label{fig:Eve_distr_ray}
    \end{subfigure}
    \caption{Loss function and data distribution for different nodes with Discrete $tanh$ function in Rayleigh wiretap channel.}
    \label{fig:loss_Q_distr_ray}
\end{figure*}


\section{Simulation and Results}

\label{results}
We implemented our experiments in TensorFlow~\cite{tensorflow}. 
In our experiments, we choose $N=96$ such that $N$ is double the number of the transmitted data subcarriers (48) in 64-pt FFT implementation of Wi-Fi standard~\cite{ieee_standered}. 
During the training phase, both the plain data and the key are random numbers generated from two different seeds, such that every block consists of different data and key combination. 
The training data size consists of $20,000$ symbols each with size $N$. 
The key to data ratio used to train the system is $0.005$. In other words, the same set of keys were repeated during the training process to ensure that the algorithm is robust and is not restricted to one-time pad only.
The batch size is $8000$. We used Adam optimizer~\cite{adam} with a learning rate of $0.001$. The number of training epoch is $4000$ for clear wiretap channel, $7000$ for Gaussian wiretap channel and $8000$ for Rayleigh wiretap channel. The three networks are trained simultaneously in each epoch such that, the weights of Eve's network are frozen while Alice and Bob update their weights and biases based on the cooperative loss function, then the weights of Alice and Bob are frozen and Eve updates her network with the updated weights and biases based on her loss function. In this work, we choose the number of discrete levels $L = 13$ for $tanh_D(x)$ activation function in Gaussian and Rayleigh channels. 
For testing phase, we use a testing data set consisting of $1000$ symbols, each of size $N$. The same key to data ratio is used in testing phase as well. However, the seeds used for data and key generation are different from those used in the training phase. For Gaussian and Rayleigh channels, the SNR range used for testing is from $0$ to $40dB$. We trained and tested the system in clear as well as AWGN and Rayleigh channels.

\subsection{Training Phase}

\subsubsection{Clear wiretap channel}

Figure~\ref{fig:Loss_Dist_clear} presents the results during the training phase in clear wiretap channel.
Loss functions of both the networks started from a high value, as shown in figure~\ref{fig:loss_clear}. After some time, Alice and Bob succeeded to infer a way to communicate securely, while Eve can not decode the confidential data. Thus the cooperative learning between Alice and Bob succeeds in beating Eve such that they can exchange the data with perfect secrecy. As a result of that, the transmitted cipher data distribution ($C$) has the shape of Gaussian distribution with zero mean as shown in figure~\ref{fig:cipher_distr_clear}. 
Thus the cipher data $C$ do not carry any statistical properties of the original plain data $P$. Hence the cipher data has the maximum uncertainty property. 
We plot the data distribution of $\hat{P}_{Bob}$ and $\hat{P}_{Eve}$ in figures~\ref{fig:Bob_distr_clear} and~\ref{fig:Eve_distr_clear} respectively. Within that distribution, we also differentiate the data points as decoded correctly or incorrectly. It is evident that distribution of $\hat{P}_{Bob}$ is not only similar to the distribution of $P$, but also the decoded bits are correct.
On the other hand, the distribution of $\hat{P}_{Eve}$ is Gaussian, yielding half of the bits in error. Hence, Eve reaches to the maximum uncertainty as the probability of error reaches the value of 0.5.

\subsubsection{Gaussian wiretap channel}
In this section, the results of using the discrete activation layer are discussed using the Gaussian wiretap channel and shown in figure~\ref{fig:Loss_Q_distr}. The number of levels chosen for the $tanh_D$ function is 13 and the training SNR is set at 25 dB. 
The loss functions, as shown in figure~\ref{fig:Loss_Q}, are similar to that noticed in clear channel. It required more number of epochs to train the two systems as there is added noise in the channel and $C$ has discrete levels, as in figure~\ref{fig:cipher_distr}. Furthermore, Eve's final loss function is lower in Gaussian channel than clear channel as the output is discretized partially sacrificing maximum uncertainty. 
$\hat{P}_{Bob}$ has only $1$s and $0$s and decodes all the bits correctly as well, as in figure~\ref{fig:Bob_distr}. In contrast, $\hat{P}_{Eve}$ has an almost uniform distribution, with higher peaks around $0$ and $1$, as in figure~\ref{fig:Eve_distr}. This leads to $20\%$ of $\hat{P}_{Eve}$ in error, which still maintains a certain level of security.

\subsubsection{Rayleigh wiretap channel}

In this section, we analyze the training phase of the system using the Rayleigh wiretap channel, as shown in figure~\ref{fig:loss_Q_distr_ray}. 
The training SNR and number of levels have been chosen to be same as the Gaussian wiretap channel. The loss function curves, as in figure~\ref{fig:Loss_Q_ray}, are similar to that of Gaussian channel, indicating that Alice and Bob succeeded in secure exchange confidential data, whereas, Eve reaches the uncertainty zone. As we introduce more noise in the channel, the cipher distribution, $C$, deviates from Gaussian distribution as noticed in earlier scenarios. 
$\hat{P}_{Bob}$ decodes all the bits correctly, as in figure~\ref{fig:Bob_distr_ray}. On the other hand, $\hat{P}_{Eve}$ has an almost uniform distribution as noticed in figure~\ref{fig:Eve_distr_ray}, which is similar to that in Gaussian channel. This yields $16\%$ of $\hat{P}_{Eve}$ in errorin the training phase.

\begin{figure}
    \centering
    \begin{subfigure}[b]{0.24\textwidth}
        \includegraphics[width=\textwidth,height=3.4cm]{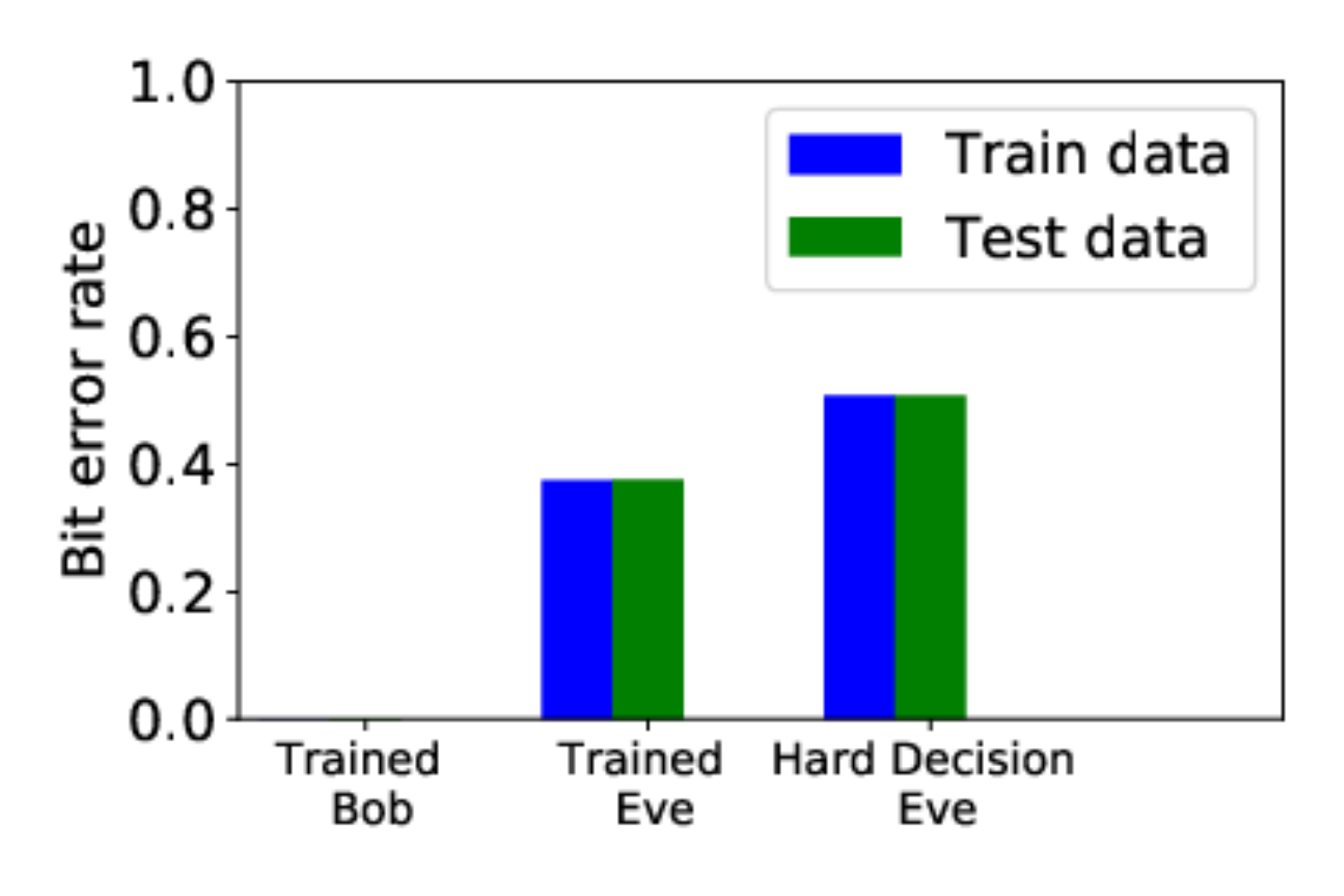}
        \caption{Clear wiretap channel.}
        \label{fig:BER_clear}
    \end{subfigure}
    \begin{subfigure}[b]{0.24\textwidth}
        \includegraphics[width=\textwidth]{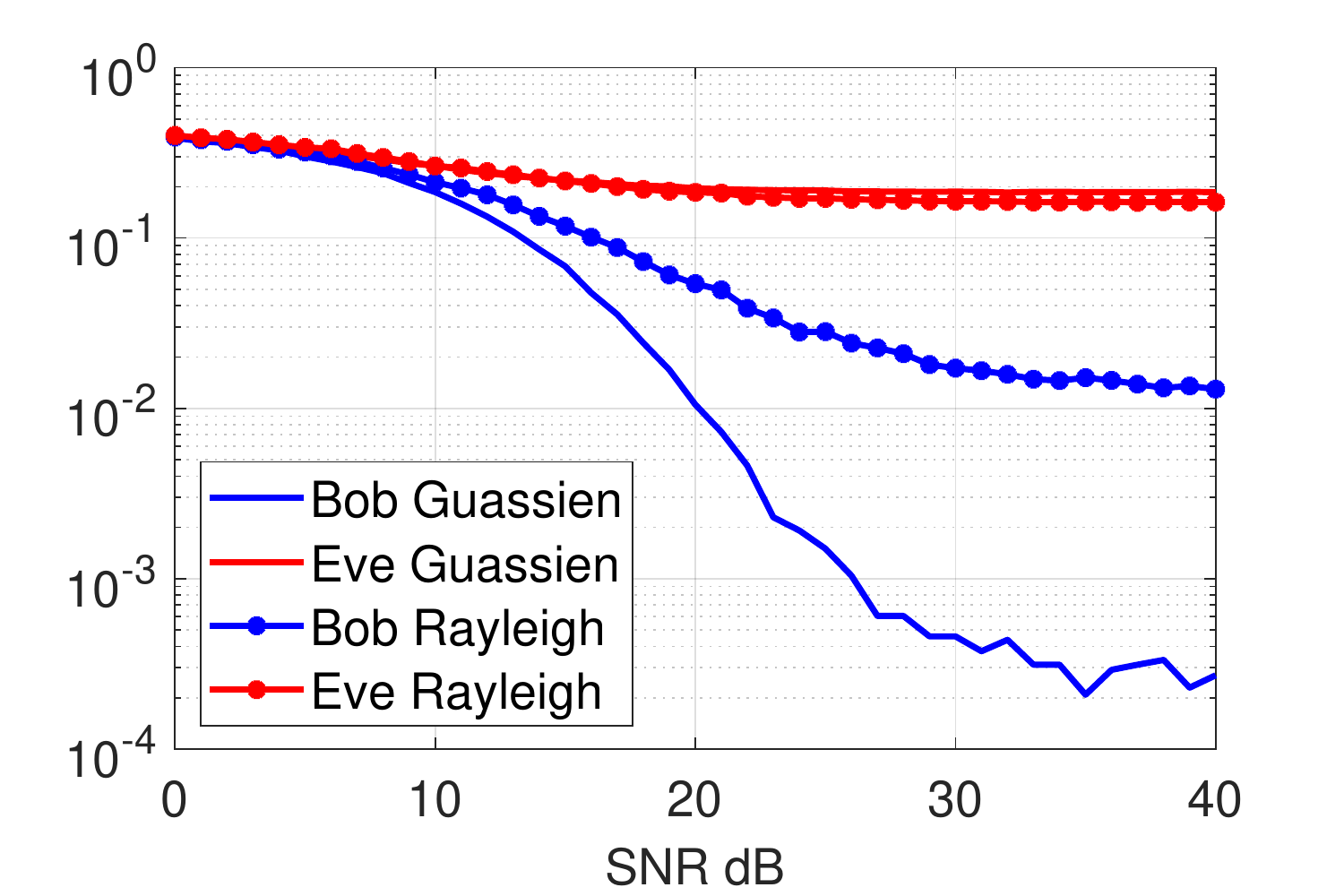}
        \caption{Noisy wiretap channels.}
        \label{fig:fig:BER_awgn}
    \end{subfigure}
    \caption{BER for Clear, Gaussian and Rayleigh channels.}
    \label{fig:BER}
\end{figure}

\subsection{Testing Phase}

Once the complete system is trained, we test the resultant networks and analyze the bit error rate (BER) in different channels, as shown in figure~\ref{fig:BER}. The source of error in BER can be of two types: a) channel imperfections as in traditional communication and b) decryption algorithm, which was unable to recover all the bits correctly. 
We define `Hard Decision Eve' as the entity, which makes a hard decision on the received cipher data ($C'$) to map to a bit value. In clear channel, $C'=C$, where the BER of `Hard Decision Eve' indicates the cross-entropy between $P$ and $C$ (i.e., $H(P/C)$). Similarly, the BER of trained Eve is a measure of cross-entropy between $P$ and $\hat{P}_{Eve}$ (i.e., $H(P/\hat{P}_{Eve})$).

In clear channel, Bob can decode both training and testing dataset correctly, as shown in figure~\ref{fig:BER_clear}. The BER of `Hard Decision Eve' is $\approx 0.5$, indicating encryption algorithm achieved the maximum value of $H(P/C)$. Trained Eve's BER is $\approx 0.4$, which is close to the maximum value of $H(P/\hat{P}_{Eve})$, which is equivalent to random guessing.
Figure~\ref{fig:fig:BER_awgn} shows the BER of trained Bob and Eve in noisy wiretap channels. Alice and Bob can securely exchange the data with a small error rate, which depends on the received SNR. On the other hand, Eve's BER remains steady at $0.2$ even at higher SNRs. Hence the learned encryption does not depend on Eve's SNR to maintain a higher value for $H(P/\hat{P}_{Eve})$.

\section{Conclusion}
\label{conc}
In this paper, we have shown that the power of neural networks can be used to learn end-to-end encrypted communication system. To improve the security of the learned encryption algorithm, we train the system in presence of an adversary to minimize the error between Alice and Bob, while maximizing the error between Alice and Eve. A discrete activation function is defined for the final modulated output to support lossy medium transmission. 
Our results indicate that a secured communication can be executed in presence of a trained or untrained Eve with the same neural network model as the trusted parties. In future, we plan to extend this work for OFDM systems and channel coding in massive MIMO wireless networks.


\bibliographystyle{IEEEtran}
\bibliography{ref.bib}

\begin{thebibliography}{10}
\providecommand{\url}[1]{#1}
\csname url@samestyle\endcsname
\providecommand{\newblock}{\relax}
\providecommand{\bibinfo}[2]{#2}
\providecommand{\BIBentrySTDinterwordspacing}{\spaceskip=0pt\relax}
\providecommand{\BIBentryALTinterwordstretchfactor}{4}
\providecommand{\BIBentryALTinterwordspacing}{\spaceskip=\fontdimen2\font plus
\BIBentryALTinterwordstretchfactor\fontdimen3\font minus
  \fontdimen4\font\relax}
\providecommand{\BIBforeignlanguage}[2]{{%
\expandafter\ifx\csname l@#1\endcsname\relax
\typeout{** WARNING: IEEEtran.bst: No hyphenation pattern has been}%
\typeout{** loaded for the language `#1'. Using the pattern for}%
\typeout{** the default language instead.}%
\else
\language=\csname l@#1\endcsname
\fi
#2}}
\providecommand{\BIBdecl}{\relax}
\BIBdecl

\bibitem{gregor2015draw}
K.~Gregor, I.~Danihelka, A.~Graves, D.~J. Rezende, and D.~Wierstra, ``Draw: A
  recurrent neural network for image generation,'' \emph{arXiv preprint
  arXiv:1502.04623}, 2015.

\bibitem{ou2007multi}
G.~Ou and Y.~L. Murphey, ``Multi-class pattern classification using neural
  networks,'' \emph{Pattern Recognition}, vol.~40, no.~1, pp. 4--18, 2007.

\bibitem{sallab2017deep}
A.~E. Sallab, M.~Abdou, E.~Perot, and S.~Yogamani, ``Deep reinforcement
  learning framework for autonomous driving,'' \emph{Electronic Imaging}, vol.
  2017, no.~19, pp. 70--76, 2017.

\bibitem{Poor19}
\BIBentryALTinterwordspacing
H.~V. Poor and R.~F. Schaefer, ``Wireless physical layer security,''
  \emph{Proceedings of the National Academy of Sciences}, vol. 114, no.~1, pp.
  19--26, 2017. [Online]. Available:
  \url{https://www.pnas.org/content/114/1/19}
\BIBentrySTDinterwordspacing

\bibitem{Survey}
J.~Hamamreh, H.~M.~Furqan, and H.~Arslan, ``Classifications and applications of
  physical layer security techniques for confidentiality: A comprehensive
  survey,'' \emph{IEEE Communications Surveys and Tutorials}, vol.~PP, pp.
  1--1, 10 2018.

\bibitem{sec_cogradio_13}
A.~G. {Fragkiadakis}, E.~Z. {Tragos}, and I.~G. {Askoxylakis}, ``A survey on
  security threats and detection techniques in cognitive radio networks,''
  \emph{IEEE Communications Surveys Tutorials}, vol.~15, no.~1, pp. 428--445,
  First 2013.

\bibitem{sec_cogradio_15}
R.~K. {Sharma} and D.~B. {Rawat}, ``Advances on security threats and
  countermeasures for cognitive radio networks: A survey,'' \emph{IEEE
  Communications Surveys Tutorials}, vol.~17, Secondquarter 2015.

\bibitem{SGD}
L.~Bottou, ``Large-scale machine learning with stochastic gradient descent,''
  in \emph{Proceedings of COMPSTAT'2010}.\hskip 1em plus 0.5em minus
  0.4em\relax Springer, 2010, pp. 177--186.

\bibitem{AE}
T.~O’Shea and J.~Hoydis, ``An introduction to deep learning for the physical
  layer,'' \emph{IEEE Transactions on Cognitive Communications and Networking},
  vol.~3, no.~4, pp. 563--575, 2017.

\bibitem{AE_GANS}
T.~J. O'Shea, T.~Roy, N.~West, and B.~C. Hilburn, ``Physical layer
  communications system design over-the-air using adversarial networks,'' in
  \emph{2018 26th European Signal Processing Conference (EUSIPCO)}.\hskip 1em
  plus 0.5em minus 0.4em\relax IEEE, 2018, pp. 529--532.

\bibitem{AE_ofdm}
A.~Felix, S.~Cammerer, S.~D{\"o}rner, J.~Hoydis, and S.~Ten~Brink,
  ``Ofdm-autoencoder for end-to-end learning of communications systems,'' in
  \emph{2018 IEEE 19th International Workshop on Signal Processing Advances in
  Wireless Communications (SPAWC)}.\hskip 1em plus 0.5em minus 0.4em\relax
  IEEE, 2018, pp. 1--5.

\bibitem{H_AE}
P.~G. Pachpande, M.~H. Khadr, H.~Hussien, H.~Elgala, and D.~Saha, ``Autoencoder
  model for ofdm-based optical wireless communication,'' in \emph{Signal
  Processing in Photonic Communications}.\hskip 1em plus 0.5em minus
  0.4em\relax Optical Society of America, 2019, pp. SpT2E--3.

\bibitem{abadi2016learning}
M.~Abadi and D.~G. Andersen, ``Learning to protect communications with
  adversarial neural cryptography,'' \emph{arXiv preprint arXiv:1610.06918},
  2016.

\bibitem{coutinho2018learning}
M.~Coutinho, R.~de~Oliveira~Albuquerque, F.~Borges, L.~Garc{\'\i}a~Villalba,
  and T.-H. Kim, ``Learning perfectly secure cryptography to protect
  communications with adversarial neural cryptography,'' \emph{Sensors},
  vol.~18, no.~5, p. 1306, 2018.

\bibitem{fritschek2019deep}
R.~Fritschek, R.~F. Schaefer, and G.~Wunder, ``Deep learning for the gaussian
  wiretap channel,'' in \emph{ICC 2019-2019 IEEE International Conference on
  Communications (ICC)}.\hskip 1em plus 0.5em minus 0.4em\relax IEEE, 2019, pp.
  1--6.

\bibitem{shannon1949communication}
C.~E. Shannon, ``Communication theory of secrecy systems,'' \emph{Bell system
  technical journal}, vol.~28, no.~4, pp. 656--715, 1949.

\bibitem{one_time}
F.~Rubin, ``One-time pad cryptography,'' \emph{Cryptologia}, vol.~20, no.~4,
  pp. 359--364, 1996.

\bibitem{shannon1948mathematical}
C.~E. Shannon, ``A mathematical theory of communication,'' \emph{Bell system
  technical journal}, vol.~27, no.~3, pp. 379--423, 1948.

\bibitem{DL_book}
I.~Goodfellow, Y.~Bengio, and A.~Courville, \emph{Deep learning}.\hskip 1em
  plus 0.5em minus 0.4em\relax MIT press, 2016.

\bibitem{Dis}
S.~Baluja, D.~Marwood, M.~Covell, and N.~Johnston, ``No multiplication? no
  floating point? no problem! training networks for efficient inference,''
  \emph{arXiv preprint arXiv:1809.09244}, 2018.

\bibitem{vanish}
R.~Pascanu, T.~Mikolov, and Y.~Bengio, ``On the difficulty of training
  recurrent neural networks,'' in \emph{International conference on machine
  learning}, 2013, pp. 1310--1318.

\bibitem{Xavier_initialization}
X.~Glorot and Y.~Bengio, ``Understanding the difficulty of training deep
  feedforward neural networks,'' in \emph{Proceedings of the thirteenth
  international conference on artificial intelligence and statistics}, 2010,
  pp. 249--256.

\bibitem{tensorflow}
M.~Abadi, P.~Barham, J.~Chen, Z.~Chen, A.~Davis, J.~Dean, M.~Devin,
  S.~Ghemawat, G.~Irving, M.~Isard \emph{et~al.}, ``Tensorflow: A system for
  large-scale machine learning,'' in \emph{12th $\{$USENIX$\}$ Symposium on
  Operating Systems Design and Implementation ($\{$OSDI$\}$ 16)}, 2016, pp.
  265--283.

\bibitem{ieee_standered}
I.~C. S. L.~S. Committee \emph{et~al.}, ``Ieee standard for information
  technology-telecommunications and information exchange between systems-local
  and metropolitan area networks-specific requirements part 11: Wireless lan
  medium access control (mac) and physical layer (phy) specifications,''
  \emph{IEEE Std 802.11\^{}}, 2007.

\bibitem{adam}
D.~P. Kingma and J.~Ba, ``Adam: A method for stochastic optimization,''
  \emph{arXiv preprint arXiv:1412.6980}, 2014.

\end{thebibliography}
\end{document}